\documentclass{jfm}

\usepackage{graphics}
\usepackage{booktabs}
\usepackage{epsfig,color,soul}
\usepackage[dvipsnames]{xcolor}
\usepackage{float,bm,dsfont}
\usepackage{natbib}
\usepackage{amsmath,amssymb,bm,gensymb} 
\usepackage{mathptmx} 
\usepackage[outdir=./]{epstopdf}
\usepackage[utf8]{inputenc}

\usepackage{array}
\newcolumntype{L}[1]{>{\raggedright\let\newline\\\arraybackslash\hspace{0pt}}m{#1}}
\newcolumntype{C}[1]{>{\centering\let\newline\\\arraybackslash\hspace{0pt}}m{#1}}
\newcolumntype{R}[1]{>{\raggedleft\let\newline\\\arraybackslash\hspace{0pt}}m{#1}}

\def\ba#1\ea{\begin{align}#1\end{align}}
\def\bsa#1#2\esa{\begin{subequations}\label{#1} \begin{align}#2\end{align} \end{subequations}}

\def\lp{\left(}\def\rp{\right)}\def\lb{\left[}\def\rb{\right]}
\def\lcb{\left\{}\def\rcb{\right\}}
\allowdisplaybreaks

\def\orange{\textcolor{orange}}
\definecolor{darkblue}{RGB}{83,0,93}

\def\Lshort{\mbox{---}}

\def\NA{\bm{\nabla}}\def\DEL{\nabla^2}\def\f{\frac}\def\p{\partial}

\def\u{{\bf{u}}}

\def\C{^{\circ} C}
\def\ba#1\ea{\begin{align}#1\end{align}}
\def\bsa#1#2\esa{\begin{subequations}\label{#1}
\begin{align}#2\end{align} \end{subequations}}

\title[Turbulent Convection in Subglacial Lakes]{Turbulent Convection in Subglacial Lakes}

\author[L.-A. Couston]{Louis-Alexandre Couston$^{1,2}$\thanks{louis.couston@ens-lyon.fr}}

\affiliation{$^1$ British Antarctic Survey, Cambridge, CB3 0ET, UK \\ $^2$ Univ Lyon, ENS de Lyon, Univ Claude Bernard, CNRS, Laboratoire de Physique, F-69342 Lyon, France}

\begin{document}
\maketitle

\begin{abstract}
Subglacial lakes are isolated, cold-temperature and high-pressure water environments hidden under ice sheets, which might host extreme microorganisms. Here, we use two-dimensional direct numerical simulations in order to investigate the characteristic temperature fluctuations and velocities in freshwater subglacial lakes as functions of the ice overburden pressure, $p_i$, the water depth, $h$, and the geothermal flux, $F$. Geothermal heating is the unique forcing mechanism as we consider a flat ice-water interface. Subglacial lakes are fully convective when $p_i$ is larger than the critical pressure $p_*\approx 2848$ dbar, but self organize into a lower convective bulk and an upper stably-stratified layer when $p_i < p_*$, because of the existence at low pressure of a density maximum at temperature $T_d$ greater than the freezing temperature $T_f$. For both high and low $p_i$, we demonstrate that the Nusselt number $Nu$ and Reynolds number $Re$ satisfy classical scaling laws provided that an effective Rayleigh number $Ra_{eff}$ is considered. We show that the convective and stably-stratified layers at low pressure are dynamically decoupled {at leading order} because plume penetration is weak and induces limited entrainment of the stable fluid. From the empirical power law equation for $Nu$ with $Ra_{eff}$, we derive two sets of closed-form expressions for the variables of interest, including the unknown bottom temperature, in terms of the problem parameters $p_i$, $h$ and $F$. The two predictions correspond to two limiting regimes obtained when the effective thermal expansion coefficient is either approximately constant or linearly proportional to the temperature difference driving the convection.

%
\end{abstract}

\section{Introduction}\label{sec:intro}

Subglacial lakes are water environments trapped between ice sheets and bedrocks \cite[][]{Siegert2001a}. Over 400 subglacial lakes have been identified in Antarctica \cite[][]{Wright2012} and about 50 have been detected in Greenland \cite[][]{Bowling2019}. Antarctica has 250 subglacial lakes that are \textit{stable}, i.e., with water trapped for millions of years and in complete isolation from Earth's climate. The remainder 150 are hydrologically \textit{active}, i.e., are connected through networks of subglacial channels and communicate via filling and discharge with the surrounding ocean \cite[][]{Smith2009}. Here, we focus on stable subglacial lakes, which are of considerable interest to astrobiology since they could host microorganisms that might have had developed novel survival strategies relevant to oceans of icy moons \cite[][]{Cockell2011}.  

Subglacial lakes are heated by Earth's geothermal flux, hence are prone to vertical convection and can experience dynamic conditions. The water circulation in stable subglacial lakes can also be driven or be affected by horizontal temperature gradients along the ice-water interface when it is tilted, due to the pressure-dependence of the freezing temperature \cite[][]{Wells2008}. The slope of the ice-water interface is typically on the order or smaller than $10^{-2}$ \cite[][]{Siegert2005}, although here we will assume for simplicity that the ice-water interface is flat. Salt concentration levels are expected to be on the order of 0.1$\%$  or less in most subglacial lakes such that the water is typically fresh \cite[][]{Siegert2001a}. A hypersaline lake has yet been recently identified in the Canadian Arctic \cite[][]{Rutishauser2018} suggesting that high salt concentrations remain possible. Subglacial lakes differ from ice-covered lakes because they typically have a much thicker ice cover and because they do not experience radiative heating \cite[][]{Ulloa2018}.

Subglacial lakes under a thick ice cover, i.e., such as Lake Vostok, which lies beneath 4 km of ice \cite[][]{Siegert2001a}, are known to be unstable to vertical convection because the thermal expansion coefficient of water, $\beta$, is always positive at high pressures. Subglacial lakes under less than about 3 km of ice, such as Lake CECs \cite[][]{Rivera2015}, may on the contrary be stable against vertical convection because $\beta<0$ at low temperatures and for pressures lower than $p_*\approx 2848$ dbar \cite[][]{Thoma2010a}. \citet[][]{Couston2020} recently proposed that the geothermal flux, which is on the order of 50 mW/m$^2$, is large enough to trigger convection in most subglacial lakes despite the nonlinearity of the equation of state. Convection typically occurs when the geothermal flux $F$ forces a bottom temperature $\overline{T}_b>T_d$ in the static state, with $T_d$ the temperature of density maximum, such that $\beta(\overline{T}_b)>0$, which is a condition met by most lakes deeper than a few meters. 

The existence of a density maximum at temperature $T_d>T_f$ with $T_f$ the freezing temperature means that low-pressure subglacial lakes self organize into a lower convective layer coupled to an overlaying stably-stratified fluid region. This two-layer dynamics has been extensively studied at atmospheric pressure, for which $T_d\approx 4\; ^{\circ}$C, both numerically \cite[][]{Lecoanet2015,Toppaladoddi2018,Wang2019} and experimentally \cite[][]{Large2014,Leard2020}. Here, using direct numerical simulations (DNS), we investigate the turbulent dynamics of freshwater environments for different ice overburden pressures, $p_i$, which enclose and include $p_*$. Thus, our results generalize the study of two-layer freshwater systems to arbitrary ice overburden pressure. 

An important point is that we consider a fixed top freezing temperature and a fixed bottom heat flux conditions, such that {our boundary conditions are different from the classical isothermal top and bottom boundary conditions considered in the canonical Rayleigh-B\'enard problem as well as by most numerical studies of mixed convective and stably-stratified fluids \cite[][]{Couston2017,Toppaladoddi2018,Wang2019}. Laboratory and numerical experiments have shown that convection driven by a bottom isothermal boundary is statistically equivalent to convection driven by a bottom fixed-flux boundary (assuming a top isothermal boundary in both cases), provided that the temperature-based Rayleigh number is the same in both experiments and the fluid is in the Oberbeck-Boussinesq regime, i.e., its properties are independent of flow velocity and temperature \cite[][]{Verzicco2008,Johnston2009}.} {We will show that the same is true for subglacial lakes, even though they are not in the Oberbeck-Boussinesq regime, provided that an effective Rayleigh number is considered. We will demonstrate that there exists two limiting behaviors of the dimensional variables with the input heat flux and water depth depending on whether the thermal expansion coefficient $\beta$ is quasi constant or linearly-varying with the temperature difference driving convection. Importantly, our results support the idea that the convective and stably-stratified layer dynamics are decoupled at leading order}, which is an hypothesis that was recently invoked in order to predict flow velocities in Antarctic subglacial lakes \cite[][]{Couston2020}. 

We organize the paper as follows. We present the equations and numerical experiments in \S\ref{sec:eqns}. We analyse the DNS results and present the theoretical predictions in \S\ref{sec:res}. We discuss the geophysical implications in \S\ref{sec:discussion} and conclude in \S\ref{sec:conc}.

\section{Problem formulation}\label{sec:eqns}

\subsection{Governing equations in dimensional form}

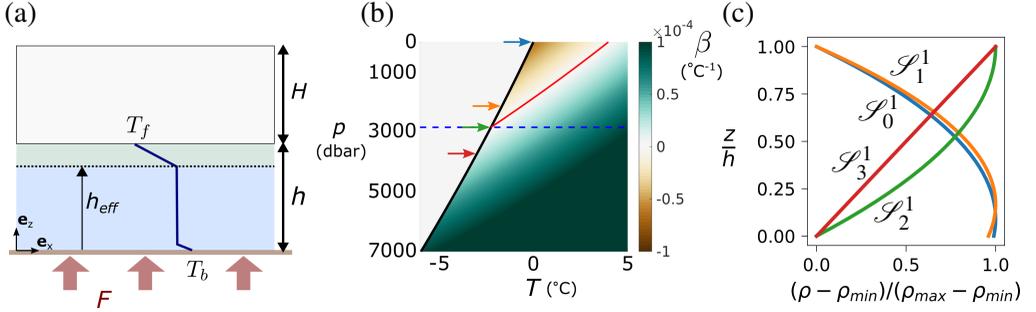
\begin{figure}
\centering
\includegraphics[width=0.99\textwidth]{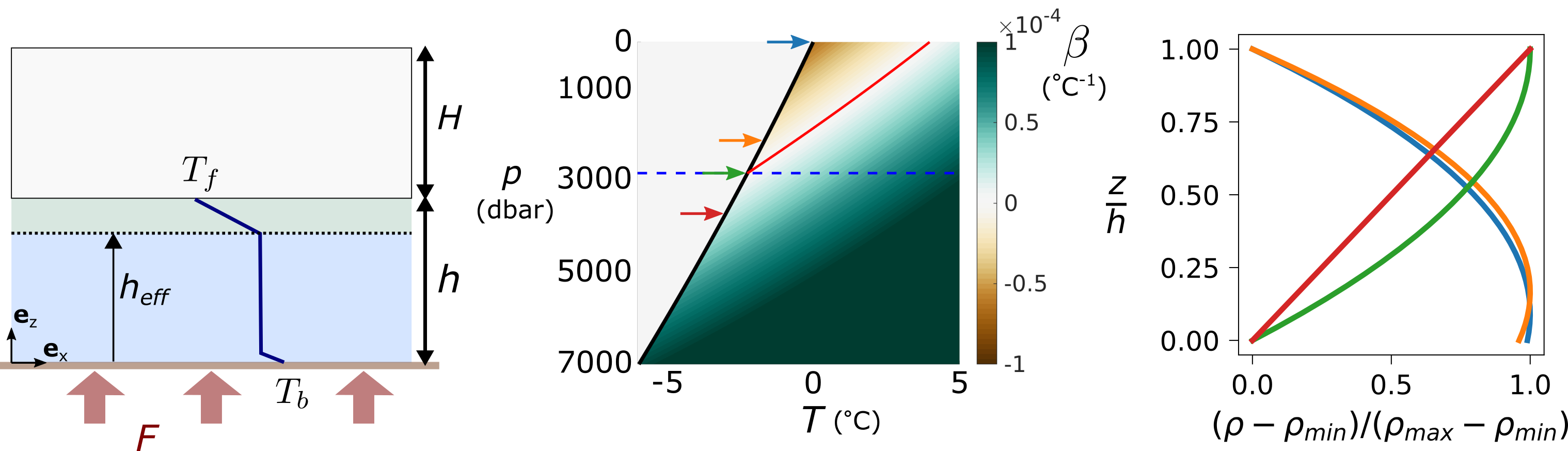}
\put(-382,107){\large{(a)}}
\put(-248,107){\large{(b)}}
\put(-112,107){\large{(c)}}
\put(-61,70){\large{$\mathcal{S}_0^1$}}
\put(-49,87){\large{$\mathcal{S}_1^1$}}
\put(-55,32){\large{$\mathcal{S}_2^1$}}
\put(-71,51){\large{$\mathcal{S}_3^1$}}
\vspace{-0.0in}\caption{(a) Problem schematic. The green shading highlights the region of the water column that stays stably stratified when $p_i \leq p_*$. (b) Thermal expansion coefficient $\beta(T,p)$. The solid black (resp. red) line shows $T_f$ (resp. $T_d$) while the dashed line shows the $p_*$ isobar. The arrows highlight the ice overburden pressures considered. (c) Density variations with depth at $t=0$ for each of the four simulation cases $\mathcal{S}_i^1$ ($i=0,1,2,3$) of the 1$^{st}$ experiment corresponding to the different ice pressures $p_i$ shown by arrows in (b).}
\label{fig1}
\end{figure}

We consider a Cartesian coordinates system $(x,y,z)$ centred on the lake's bottom boundary with $\bold{e}_z$ the upward-pointing unit vector of the $z$ axis, i.e., opposite to gravity, and we denote $H$ the ice thickness and $h$ the lake water depth (cf. figure \ref{fig1}(a)). For computational expediency we restrict our attention to two-dimensional motions, i.e., we assume $y$ invariance and neglect rotation. Here, as in most liquids, compressibility effects are weak and density fluctuations with temperature and pressure are small compared to the reference density $\rho_0=999$ kg/m$^3$. As a result, the evolution of the lake's velocity $\u$ and temperature $T$ is well approximated by the Navier-Stokes equations in the Boussinesq approximation and the incompressible energy equation, i.e., such that
\bsa{eq:a1}\label{eq:a11}
& \p_t \u - \nu\DEL\u + \NA (p/\rho_0) = -  \lp\u\cdot\NA\rp\u    - (\rho/\rho_0) g\bold{e}_z, \\ \label{eq:a12}
&\NA\cdot \u = 0,  \\ \label{eq:a13}
& \p_t T - \kappa \DEL T = - \lp\u\cdot\NA\rp T,
\esa
where $p$ is the pressure, $\rho$ is the density, $\p_t$ denotes time derivative and $\nabla$ is the gradient operator. The physical parameters in \eqref{eq:a1} are the kinematic viscosity $\nu$, the reference density $\rho_0$, the gravitational acceleration $g$ and {the thermal diffusivity $\kappa$} (cf. table \ref{tab:phy}). For the boundary conditions, we consider
\ba{}\label{eq:bcs}
\u(z=0)=\u(z=h)=\bold{0}, \quad \p_zT(z=0)=-F/k, \quad T(z=h)=T_f(p_i),
\ea
i.e., we assume no-slip, fixed heat flux $F$ on the bottom boundary {with $k$ the thermal conductivity}, and we set the temperature at the top of the lake equal to the temperature of freezing, $T_f$, which varies with the ice overburden pressure $p_i$.

We approximate the equation of state for the density of freshwater as a function of the lake pressure $p\geq p_i$ and temperature $T\geq T_f(p_i)$ using the bivariate polynomial 
\ba{}\label{eq:eos}
\rho(p,T)=\rho_0 + \rho_1(p) + C(p)\lb T-T_d(p)\rb^2,
\ea
where $T_d$ is the temperature of maximum density, i.e., such that $(\p \rho/\p T)|_p(T=T_d)=0$. We obtain the (quadratic) polynomial expressions for $\rho$ (through $\rho_1$ and $C$), $T_d$ and $T_f$ as functions of pressure by minimizing their L$^2$ relative error norm compared to the exact thermodynamic values $\rho^e$, $T_d^e$ and $T_f^e$ (superscript $^e$ denoting exact values) computed using  TEOS-10 \cite[][]{McDougall2011}. The polynomial approximations for $\rho$, $T_d$ and $T_f$ (provided in table \ref{tab:phy}) result in errors smaller than 0.1 $g/m^3$ and 0.002 $\C$ for $p,p_i\in[0,10000]$ dbar and $T\in[T_f,T_f+15 \; ^{\circ}$C]. Figure \ref{fig1}(b) shows the pressure-dependence of $T_f$ (solid black line) and $T_d$ (red line). Both $T_f$ and $T_d$ decrease with increasing pressure, but $T_d > T_f$, i.e., such that the water is densest at a non-freezing temperature, only for $p<p_*= 2848.5$ dbar, which we call the critical ice overburden pressure (dashed blue line). The form of the equation of state \eqref{eq:eos} highlights that the density can be non-monotonic with temperature and exhibit a maximum at $T=T_d$ within the water column provided that $T_d(p)>T_f(p_i)$. This condition requires $p_i < p_*$ and is most likely to be satisfied at the top of subglacial lakes since $p\geq p_i$ increases with depth by hydrostasy and $T_d$ decreases with $p$. The thermal expansion coefficient is
\ba{}\label{eq:thexp}
\beta = -\f{1}{\rho_0}\left.{\f{\partial \rho}{\partial T}}\right|_{p} = -\f{2C(p)[T-T_d(p)]}{\rho_0}.
\ea
Figure \ref{fig1}(b) clearly shows that $\beta>0$ for all temperatures when $p>p_*$, while $\beta$ can change sign with temperature for $p<p_*$, i.e., when the temperature of maximum density exceeds the freezing temperature. The critical ice-cover thickness associated with $p_*$ is $H_*=10^4p_*/(\rho_ig)=3166$ m ($p_*$ in dbar) assuming a mean ice density $\rho_i=917$ kg/m$^3$.

The density $\rho$ and the temperature of maximum density $T_d$ are functions of the full pressure $p$ (cf. table \ref{tab:phy}). {However, for simplicity, here we will substitute $\rho(p,T)$ and $T_d(p)$ with $\rho(p_i,T)$ and $T_d(p_i)$ in the governing equations, i.e., such that $\rho$ and $T_d$ depend on the ice overbuden pressure only. This approximation is legitimate for lakes that are not too deep, i.e., such that hydrostatic pressure variations are weak and considering $\rho(p,T)\approx \rho(p_i,T)$ and $T_d(p)\approx T_d(p_i)$ does not impact significantly buoyancy effects. All lakes considered in this work are shallow, i.e., the water depth does not exceed 8 meters, such that the approximation is valid. In particular, a simulation of a lake with a maximum depth of 8 meters yields almost identical results whether we make or relax the assumption $\rho(p,T)\approx \rho(p_i,T)$ and $T_d(p)\approx T_d(p_i)$ (cf. appendix \ref{sec:appA}). Note that approximating $\rho(p,T)\approx \rho(p_i,T)$ implies approximating  $\beta(p,T)\approx \beta(p_i,T)$ too.}
\begin{table}\centering \begin{tabular}{L{5cm}L{10cm}}
Physical parameters & Polynomial expressions \\ 
$\rho_0 = 9.9999\; 10^{2} \; \f{kg}{m^3}$ & $T_f(p_i) = 4.7184\; 10^{-3} -7.4584\; 10^{-4}p_i -1.4999\; 10^{-8}p_i^2$  \\
$g = 9.81\; \f{m}{s^2}$ & $T_d(p) = 3.9795-2.0059\; 10^{-3}p-6.2514\; 10^{-8}p^2$  \\ 
$\nu = 1.70 \; 10^{-6} \; \f{m^2}{s}$ & $\rho_1(p) = 4.9195\; 10^{-3}  p - 1.4372\; 10^{-8} p^2 $ \\ 
$\kappa =  1.33\; 10^{-7} \; \f{m^2}{s}$ & $C(p) = - 7.0785\; 10^{-3} + 1.8217\; 10^{-7}p + 4.2679\; 10^{-12}p^2$ \\
$k = 0.56 \; \f{W}{m\C}$ & \\
\hline
\end{tabular}\vspace{-0.in}\caption{Physical parameters and polynomial approximations for $T_f$, $T_d$, $\rho_1$ and $C$ with temperatures in $\C$, pressures in $dbar$, densities in $kg/m^3$ and $C$ in $kg/m^3/\C^2$.}\label{tab:phy}\end{table}

{Our study of natural convection in subglacial lakes is fundamentally a study of non-Oberbeck-Boussinesq (NOB) effects in thermal convection due to a temperature-dependent thermal expansion coefficient \eqref{eq:thexp}.} Previous works on NOB effects due to a temperature-dependent thermal expansion coefficient that can change sign include \cite[][]{Couston2017,Toppaladoddi2018,Wang2019}. \cite{Toppaladoddi2018} and \cite{Wang2019} considered the equation of state for water at constant atmospheric pressure, i.e., such that their range of $\beta<0$ was fixed, whereas it varies with pressure in our case (see, e.g., figure \ref{fig1}(b)). \cite{Couston2017} used a piecewise-linear equation of state and a variable \textit{stiffness} parameter, which allowed them to consider different ranges for $\beta<0$. {Our work is different from \cite{Couston2017} because (i) we consider the full equation of state for water rather than an artificial equation of state and (ii) the bottom boundary conditions is fixed heat flux in our work rather than fixed temperature, which we will show is an important point when $\beta$ varies with $T$.} {The dependence of viscosity $\nu$ and thermal diffusivity $\kappa$ with temperature are two other well-known NOB effects that can lead to noticeable deviations of thermal convection, including a top-down asymmetry, from the classical Rayleigh-B\'enard experiment \cite[][]{Ahlers2006,Sugiyama2009}. Nevertheless, here we take $\nu$ and $\kappa$ as constants since their relative variations do not exceed 50\% over the range of $(p,T)$ considered, i.e., 0 dbar$<p<10^4$ dbar and -5 $^{\circ}$C$<T<$5 $^{\circ}$C \cite[][]{Forst2000,Huber2012} and are not expected to have an effect on the lakes' dynamics as important as the variations of the thermal expansion coefficient. Future works might consider relaxing this assumption.}

\subsection{Governing equations in dimensionless form}\label{sec:dless}

We use the water depth $h$ as characteristic length scale, the diffusive time  $\tau_{\kappa}=h^2/\kappa$ as time scale, the velocity $u_{\kappa}=h/\tau_{\kappa}$ as velocity scale and the pressure $p_{\kappa}=\rho_0u_{\kappa}^2$ as pressure scale in order to identify dimensionless control parameters and non-dimensionalize the governing equations, which we recall are equations \eqref{eq:a1}, \eqref{eq:bcs} and \eqref{eq:eos} with $\rho(p_i,T)$ substituted for $\rho(p,T)$. The temperature drop between the lake's top and bottom boundaries in the turbulent regime is unknown. However, we can use ${\Delta} = Fh/k$ as temperature scale, which is the temperature drop across the lake's depth of the diffusive base state, which we denote by overbars and is given by $\overline{\u}=\bold{0}$, $\overline{T}=T_f+{\Delta}(1-z/h)$ and hydrostatic pressure $\overline{p}=p_i+\int_z^h \overline{\rho}gdz'$. We use $T_f(p_i)$ as reference temperature and $p_i+[\rho_0+\rho_1(p_i)]g(h-z)$ as pressure gauge, i.e., such that we remove the leading-order mean buoyancy and hydrostatic pressure terms, which balance each other, in the governing equations. {The dimensionless variables, which we denote by tildes, are given by
\ba{}\label{eq:02}
(x,z)=h(\widetilde{x},\widetilde{z}), \; t=\tau_{\kappa}\widetilde{t}, \; u=u_{\kappa}\widetilde{u}, \; p = p_i+[\rho_0+\rho_1(p_i)]g(h-z) + p_{\tau}\widetilde{p}, \; T = T_f + {\Delta} \widetilde{T}. 
\ea
}Substituting \eqref{eq:02} into \eqref{eq:a1} and \eqref{eq:bcs} combined with \eqref{eq:eos} (with $p_i$ replacing $p$ in the expression for $\rho$), yields a set of dimensionless equations and boundary conditions, which we write as
%
\bsa{eq:a2}\label{eq:a21}
& \p_{\tilde{t}} \widetilde{\u} - Pr\widetilde{\nabla}^2\widetilde{\u} + \widetilde{\NA} \widetilde{p} = -  \lp\widetilde{\u}\cdot\widetilde{\NA}\rp\widetilde{\u}    + Pr\overline{Ra}_F\f{\lp 1+\overline{S}\rp}{2}\lb \widetilde{T}-  \f{\overline{S}}{\lp 1+\overline{S}\rp}  \rb^2\bold{e}_z, \\ \label{eq:a22}
& \p_{\tilde{t}} \widetilde{T} - \widetilde{\nabla}^2 \widetilde{T} = - \lp\widetilde{\u}\cdot\widetilde{\NA}\rp \widetilde{T}, \\ \label{eq:a23}
&\widetilde{\NA}\cdot \widetilde{\u} = 0, \\
&\widetilde{\u}(\widetilde{z}=0)=\widetilde{\u}(\widetilde{z}=1)=\bold{0}, \quad \p_{\tilde{z}}\widetilde{T}(\widetilde{z}=0)=-1, \quad \widetilde{T}(\widetilde{z}=1)=0, 
\esa
{with 
\ba{}\label{eq:cpar}
Pr=\f{\nu}{\kappa}, \quad \overline{Ra}_F=\f{gh^4F\overline{\beta}_b}{k\nu\kappa}, \quad \overline{S}=\f{T_d-T_f}{\overline{T}_b-T_d},
\ea
the control parameters}, and {where
\ba{}\label{eq:diffprof}
&\overline{\beta}_b=-2C(\overline{T}_b-T_d)/\rho_0, \\\label{eq:diffTb}
&\overline{T}_b=T_f+{\Delta},
\ea
are the bottom thermal expansion coefficient and the bottom temperature of the diffusive base state, respectively.} 

The control parameters \eqref{eq:cpar} are the (constant) Prandtl number $Pr=12.8$, the base-state flux Rayleigh number $\overline{Ra}_F$, which is based on the heat flux $F$ and the bottom thermal expansion coefficient of the diffusive base state $\overline{\beta}_b$ and the base-state stiffness number $\overline{S}$, which compares the thermal expansion coefficient at the top of the lake, i.e., at temperature $T_f$, to the thermal expansion coefficient at the bottom in the diffusive base state, i.e., at temperature $\overline{T}_b$. {The base-state flux Rayleigh number $\overline{Ra}_F$ is positive and convection is possible if $\overline{\beta}_b>0$, i.e., if the bottom temperature of the diffusive base state exceeds the temperature of maximum density. This condition is satisfied provided that the heat flux exceeds a minimum heat flux for fixed $p_i$ and $h$, or, equivalently, the water depth exceeds a minimum water depth for fixed $p_i$ and $F$, which we call the threshold heat flux and threshold water depth, respectively, and define as
\ba{}\label{eq:threshold}
F_t = \text{max}\lcb \f{k\lb T_d(p_i)-T_f(p_i) \rb}{h},0 \rcb , \quad h_t = \text{max}\lcb \f{k\lb T_d(p_i)-T_f(p_i) \rb}{F}, 0 \rcb , 
\ea
i.e., such that $F_t>0$ and $h_t>0$ if $p_i<p_*$ and $F_t=h_t=0$ if $p_i\geq p_*$.  When $Fh>k[T_d(p_i)-T_f(p_i)]$, $\overline{Ra}_F$ increases monotonically with $p_i$, $h$ and $F$. The base-state stiffness number $\overline{S}$ can take any value in $[-\infty,+\infty]$. Specifically, $\overline{S}\leq -1$ when the lake is fully stable, i.e., $F<F_t$; $-1<\overline{S}\leq 0$ when the lake is fully unstable, i.e., $p_i>p_*$; and, $\overline{S}>0$ when the lake is partially convective, i.e., $T_f<T_d<\overline{T}_b$, and self organizes into a stably-stratified upper layer and a convective bottom layer. When $\overline{S}>0$, we might expect that larger stiffness parameters $\overline{S}$ correlate with stronger resistance of the top stable layer to overshooting convective motions \cite[see, e.g.,][]{Couston2017}. When the lake is fully unstable, i.e., $-1< \overline{S} \leq 0$, $\overline{S}\approx -1$ indicates that the base-state thermal expansion coefficient is almost depth invariant, while $\overline{S}\approx 0$ indicates that it strongly varies with depth.}

\subsection{Numerical experiments}\label{sec:numexp}

{Since we impose the heat flux rather than the temperature on the bottom boundary of the lake, we expect the mean bottom temperature $\langle T_b \rangle=\langle T(z=0) \rangle$, with $\langle \cdot \rangle$ denoting the horizontal and temporal average, to become smaller than $\overline{T}_b$ as convection sets in. This means that the control parameters $\overline{Ra}_F$ and $\overline{S}$, while fully prescribing the system, do not provide an effective measure of buoyancy forcing compared to dissipation or of density stratification in the stable layer when the flow is turbulent and at statistical steady state. As a result, our goals are to:
\begin{enumerate}
\item investigate the variations of the bottom temperature $\langle T_b \rangle$ at statistical steady state with the problem parameters, 
\item define an effective Rayleigh number $Ra_{eff}$ based on $\langle T_b \rangle$, which can be used to predict the characteristic Reynolds number $Re$ and Nusselt number $Nu$ of the convective layer,
\item investigate the influence of the stable layer on the convective dynamics through the use of, e.g., an effective stiffness parameter $S_{eff}$.
\end{enumerate}
We note that many studies have investigated how the classical scalings of Rayleigh-B\'enard convection (between isothermal plates) change when changing the boundary conditions (e.g., fixed heat flux) or considering NOB effects \cite[][]{Chilla2012}. However, previous studies considering the effect of a fixed-flux boundary condition have been limited to the Oberbeck-Boussinesq regime \cite[][]{Verzicco2008,Johnston2009}, while those exploring the effect of a density maximum have used isothermal boundaries \cite[][]{Couston2017,Toppaladoddi2018,Wang2019,Leard2020}. Thus, we expect that the analysis presented in this paper can provide new fundamental results on non-classical Rayleigh-B\'enard convection while being also useful to the study of subglacial lakes.}

\begin{figure}
\centering
\includegraphics[width=1\textwidth]{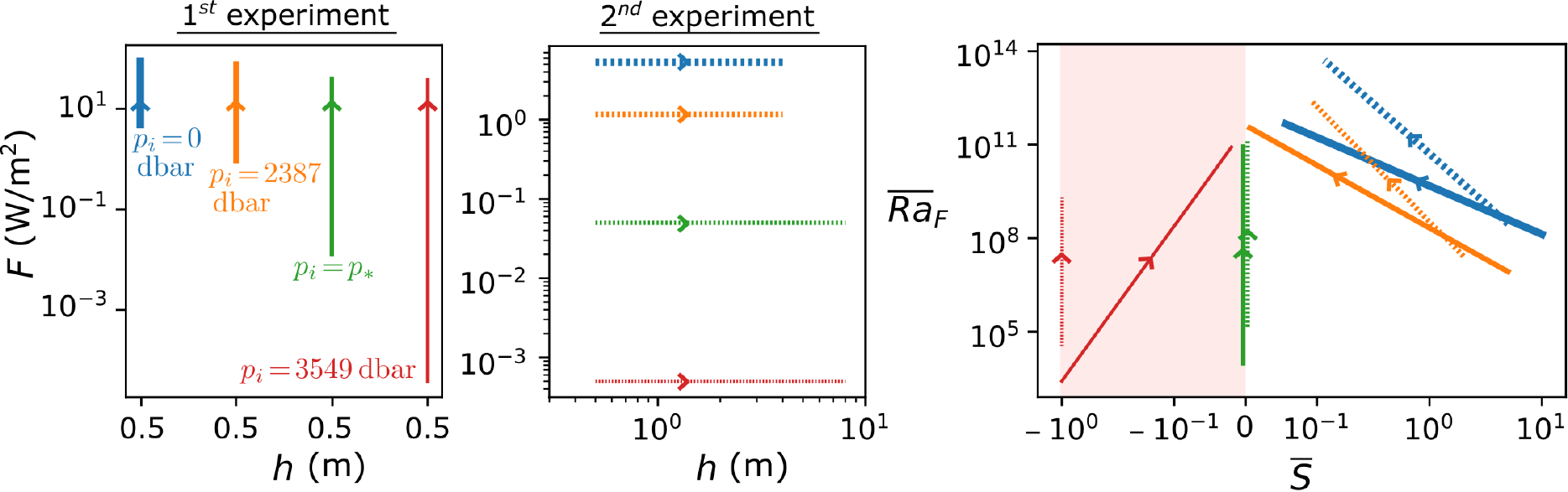}
\put(-370,115){\large{(a)}}
\put(-268,115){\large{(b)}}
\put(-160,115){\large{(c)}}
\vspace{-0.05in}\caption{Graphical illustration of our two sets of numerical experiments in physical space with (a) $h=0.5$ m fixed in the first experiment and (b) $F$ (in W/m$^2$) fixed in the second (cf. details in table \ref{tab:sims}). {Thinner lines correspond to larger ice overburden pressure.}  (c) Corresponding coverage in dimensionless space $(\overline{S},\overline{Ra}_F)$. The red shading highlights the region of fully-convective lakes.}
\label{fig2}
\end{figure}

{Although parametric fluid dynamics studies are usually optimally designed when sweeping through parameters in dimensionless space, here we explore the effect of the parameters on the flow dynamics by sweeping through the physical space $(p_i,F,h)$ rather than through the control parameter space $(\overline{Ra}_F,\overline{S})$. The reason is that we are interested in the variations and predictions of the bottom temperature $\langle T_b \rangle$ and flow velocities in the convective layer in terms of a specific range of geophysically-relevant, or laboratory-relevant, lake parameters $p_i$, $F$ and $h$, which is difficult to cover with an exploration in $(\overline{Ra}_F,\overline{S})$ space due to the nonlinear relationships between $\overline{Ra}_F$, $\overline{S}$ and $F$, $h$ and $p_i$. Thus, we conduct our investigation of subglacial lake dynamics by sweeping through lines of constant heat flux and lines of constant water depth in physical space $(F,h)$ while considering ice pressures both above and below the critical ice pressure $p_*$, which separates the fully-convective regime from the partially-convective one (figure \ref{fig2}).}

We consider subglacial lakes under 4 different ice overburden pressures, i.e., $p_i=0$, 2387, 2848.5 and 3549 dbar, corresponding to ice thicknesses $H=0$ (infinitesimally small ice layer), $H=2653$ m \cite[relevant for subglacial lake CECs, cf.][]{Rivera2015}, $H=H_*$ and $H=3945$ m \cite[relevant for subglacial lake Vostok, cf.][]{Siegert2001a}, respectively. For each ice pressure considered we investigate the temperature variations in the lake and the root-mean-square velocity as functions of the geothermal flux $F$ and water depth $h$. We use two sets of experiments. {First we focus on the case $h=0.5$ m and increase $F$ in successive stages. We denote the corresponding simulation cases $\mathcal{S}_i^1$, with $i=0..3$ increasing as $p_i$ increases (four multi-stage simulations), i.e., such that e.g. $\mathcal{S}_0^1$ corresponds to the simulation with $h=0.5$ m, $p_i=0$ dbar and $F$ increasing in successive stages. We pick $h=0.5$ m, which is a relatively standard height for water containers, such that the first set of simulations may be compared with future laboratory experiments provided that water can be pressurized.} Second, we fix $F$ and increase $h$ in stages. We denote these simulations $\mathcal{S}_i^2$, again with $i=0..3$ increasing as $p_i$ increases (cf. table \ref{tab:sims}). Each stage of a simulation lasts one diffusive thermal time such that the results, averaged over the second half of a stage, describe the system at statistical steady state. Figure \ref{fig2} highlights the physical parameter space covered by the numerical simulations as well as the corresponding coverage in dimensionless space.

{For each simulation we first compute the threshold heat flux $F_t$ if $h$ is fixed (1st set of experiments) or the threshold water depth $h_t$ if $F$ is fixed (2nd set of experiments) using equation \eqref{eq:threshold}. Then, we evaluate the critical heat flux $F_c$ (resp. critical water depth $h_c$), which is required for the destabilizing buoyancy force to overcome viscous dissipation and thermal diffusion in equations \eqref{eq:a2} for the 1st (resp. 2nd) set of experiments. For the calculation of $F_c$ and $h_c$ we use the eigentools package\footnote{https://github.com/jsoishi/eigentools} in Python, which is based on the eigenvalue-solver capability of the open-source pseudo-spectral code Dedalus \cite[][]{Burns2020}. Necessarily, $F_c > F_t$ and $h_c>h_t$. We report $F_t$ and $F_c$, as well as $h_t$ and $h_c$, and the range of supercritical heat fluxes and water depths considered for each simulation case in table \ref{tab:sims}. At $t=0$, we initialize the system with no velocities and a conductive (linear) temperature profile superimposed with small-amplitude white noise.} The corresponding mean density profiles are shown in figure \ref{fig1}(c) for the four simulations of the first experiment. For $\mathcal{S}_0^1$ and $\mathcal{S}_1^1$, for which $p_i<p_*$, the density increases with height, hence is convectively unstable, in a lower subregion of the water column but decreases with height, hence is stably stratified, above. For $\mathcal{S}_2^1$ and $\mathcal{S}_3^1$ the density always increases with height such that the full water column is unstable to convection. For $\mathcal{S}_2^1$, $p_i=p_*$ and $T_f=T_d$ such that $\beta=0$ at $z=h$, which is why $\p_z\rho=0$ at the top boundary.

\begin{table}\centering 
\begin{tabular}{C{0.9cm}C{0.7cm}C{0.6cm}C{0.4cm}C{0.9cm}C{1.5cm}C{1.8cm}C{0.2cm}C{0.6cm}C{0.8cm}C{0.7cm}C{0.7cm}C{1.1cm}}
& & \multicolumn{5}{c}{1$^{st}$ set of experiments ($h$ fixed)} & & \multicolumn{5}{c}{2$^{nd}$ set of experiments ($F$ fixed)}\\
\cmidrule{3-7}\cmidrule{9-13}
$p_i$ & $H$ & s.n. & $h$ & $F_t$ & $F_c$ & $F$ &  & s.n. & $F$ & $h_t$  & $h_c$  & $h$  \\ 
0      & 0     & $\mathcal{S}_0^1$ & 0.5 & 4.452 & 4.711           & $1.1F_t-20F_t$ &  & $\mathcal{S}_0^2$ & 5.34 & 0.42  & 0.44  & $0.5-2$ \\
2387   & 2653  & $\mathcal{S}_1^1$ & 0.5 & 0.781 & 0.876           & $1.2F_t-100F_t$ &  & $\mathcal{S}_1^2$ & 1.17 & 0.33  & 0.38  & $0.5-4$ \\
2848.5 & 3166  & $\mathcal{S}_2^1$ & 0.5 & 0     & 6.515 $10^{-3}$ & $2F_c-6\;10^3F_c$ &  & $\mathcal{S}_2^2$ & 0.05 & 0  & 0.22 & $0.5-8$ \\
3549   & 3945  & $\mathcal{S}_3^1$ & 0.5 & 0     & 1.912 $10^{-5}$ & $2F_c-2\;10^6F_c$ &  & $\mathcal{S}_3^2$ & 0.005 & 0  & 0.12 & $0.5-8$ \\
\hline
\end{tabular}\vspace{-0.in}\caption{Dimensional parameters for the two sets of numerical experiments, which consider 4 distinct ice overburden pressures $p_i$ (in dbar) each and either a broad range of geothermal fluxes (7th column) or a broad range of water depths (last column). Ice thickness $H$ and water depths $h$, $h_c$ and $h_t$ are in meters and fluxes $F$, $F_c$ and $F_t$ are in W/m$^2$; s.n. means simulation name. Note that figure \ref{fig2} provides a graphical illustration of the dimensional and dimensionless parameter spaces explored.}\label{tab:sims}\end{table}

We solve equations \eqref{eq:a2} with the open-source pseudo-spectral code Dedalus \cite[][]{Burns2020}. We assume that the $x$ direction is periodic and has dimensional length $L_x=4h$. {We recall that we use no-slip boundary conditions, an isothermal top boundary and a fixed heat flux bottom boundary. The horizontally-averaged dynamic pressure, i.e., in excess of the hydrostatic pressure, is set to 0 at the top boundary. We use a Fourier basis with  $n_x=512$ modes in the $x$ direction and a Chebyshev basis with $n_z=256$ in the $z$ direction before dealiasing for the most turbulent simulations. For the least turbulent simulations with a stable layer, i.e., $\mathcal{S}_0^1$, $\mathcal{S}_1^1$, $\mathcal{S}_0^2$, $\mathcal{S}_1^2$ with $F\leq 4F_t$, $F\leq 20F_t$, $h\leq 0.8$ m and $h\leq 1.3$ m, respectively, we decrease $n_x$ to 256. For the least turbulent fully-convective simulations, i.e., $\mathcal{S}_2^1$, $\mathcal{S}_3^1$, $\mathcal{S}_2^2$, $\mathcal{S}_3^2$ with $F\leq 1250F_c$, $F\leq 2\times 10^5F_c$, $h\leq 4$ m and $h\leq 2$ m, respectively, we decrease both $n_x$ to 256 and $n_z$ to 128. We use a second-order two-step Runge-Kutta method for time integration and a CFL between 0.2 and 0.4, with the lower CFL used for the most turbulent simulations.}

\section{Results}\label{sec:res}

\subsection{General flow features}\label{sec:gen}

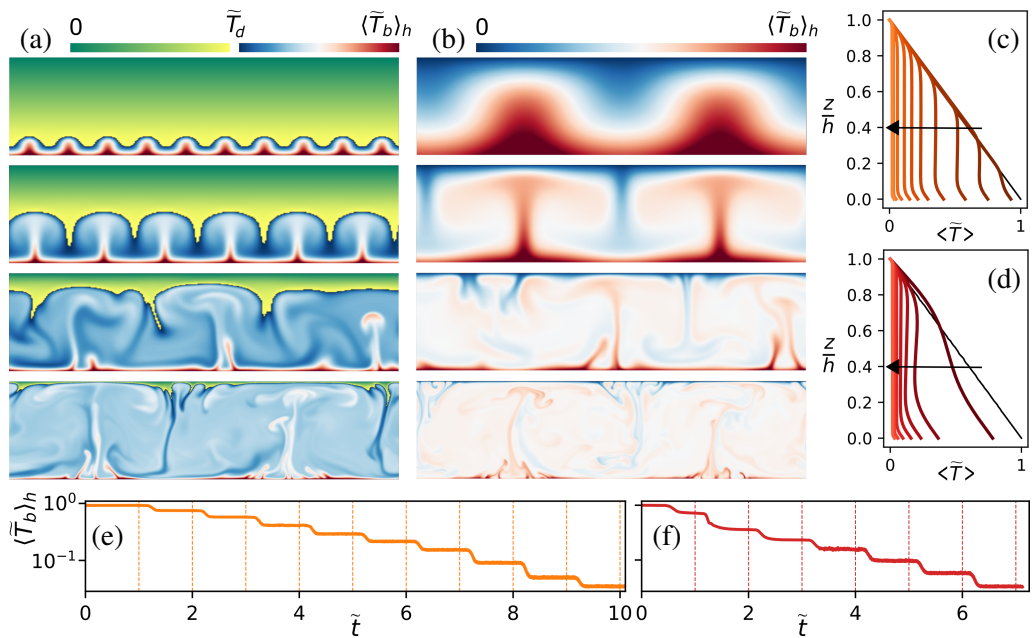
\begin{figure}
\centering
\includegraphics[width=1\textwidth]{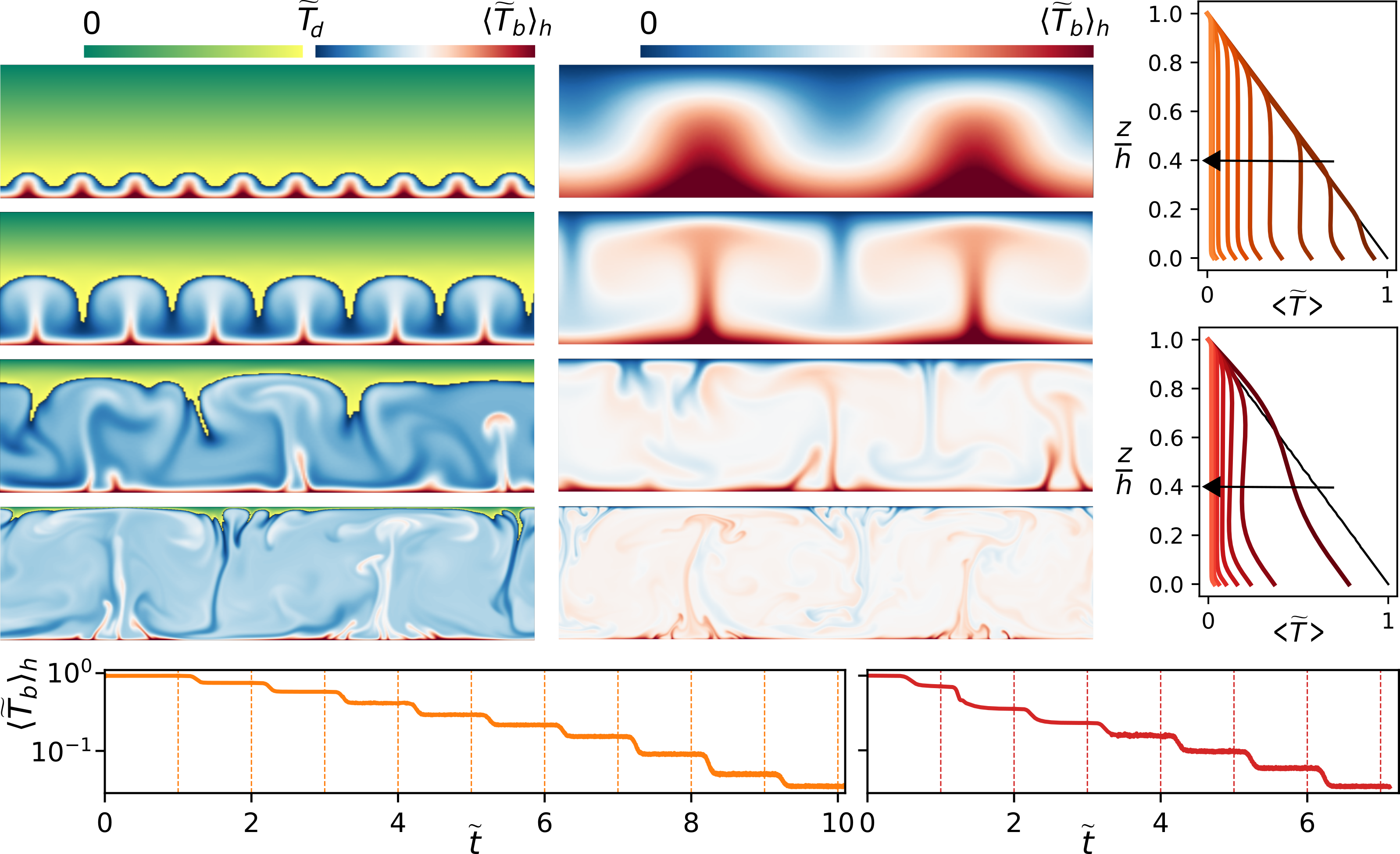}
\put(-380,220){\large{(a)}}
\put(-225,220){\large{(b)}}
\put(-16,220){\large{(c)}}
\put(-16,130){\large{(d)}}
\put(-353,34){\large{(e)}}
\put(-142,34){\large{(f)}}
\vspace{-0.in}\caption{(a) Snapshots of dimensionless temperature $\widetilde{T}$ for simulation $\mathcal{S}_1^1$ with $F$ increasing from top to bottom. $\widetilde{T}_d$  is the dimensionless temperature of maximum density whereas $\langle\widetilde{T}_b\rangle_h$ is the dimensionless horizontally-averaged (but time-dependent) bottom temperature. (b) Same as (a) but for $\mathcal{S}_3^1$. (c) Dimensionless time- and horizontally-averaged temperature profiles $\langle \widetilde{T} \rangle$ with depth at different stages (i.e., different $F$) for $\mathcal{S}_1^1$. (d) Same as (c) but for $\mathcal{S}_3^1$. The line colors go from dark to light as $F$ increases from small to large values {(lines shifting from right to left as shown by the black arrows)}. The thin black lines show the conductive profiles at $\widetilde{t}=0$. (e) and (f) show the time evolution of $\langle  \widetilde{T}_b\rangle_h$ for $\mathcal{S}_1^1$ and $\mathcal{S}_3^1$, respectively. {The vertical dashed lines highlight the times $\widetilde{t}=i$ ($i=1,2,3...$) when the control parameter ($F$ or $h$) starts increasing (smoothly) and the simulation stage changes, with a new statistical steady state reached before $\widetilde{t}=i+0.5$.}}
\label{fig3}
\end{figure}

The flow dynamics in a subglacial lake experiencing a dimensionless thermal expansion coefficient $\widetilde{\beta}=\beta/\overline{\beta}_b$ that changes sign within the water column (i.e., if $\widetilde{T}_d>\widetilde{T}_f=0$) is qualitatively different from the flow dynamics in a subglacial lake with $\widetilde{\beta}$ positive throughout. Note that tildes denote dimensionless variables and that we will normalize all thermal expansion variables by $\overline{\beta}_b$ (cf. equation \eqref{eq:diffprof}). Figures \ref{fig3}(a) and \ref{fig3}(b) show several snapshots of the dimensionless temperature field $\widetilde{T}$ for simulation $\mathcal{S}_{1}^1$, for which $\widetilde{\beta}$ changes sign inside the water column, and simulation $\mathcal{S}_{3}^1$, for which $\widetilde{\beta}>0$ everywhere. In figure \ref{fig3}(a), the water column is only partially unstable to convection. Convective motions (associated with $\widetilde{\beta}>0$), which are shown with the red-to-blue colormap, coexist with a stably-stratified layer (where $\widetilde{\beta}<0$), which is shown by the yellow-to-green colormap. As $F$ increases (from top to bottom), the system transitions from a stationary laminar state to a turbulent state, and, at the same time, the bottom convective layer grows while the top stable layer shrinks. In figure \ref{fig3}(b), $\widetilde{\beta}>0$ everywhere, such that the water column is  convecting over the full depth and the flow dynamics is qualitatively similar to classical Rayleigh-B{\' e}nard convection. 

We show in figures \ref{fig3}(c) and \ref{fig3}(d) the vertical profiles of the time-  and horizontally-averaged dimensionless temperature $\langle \widetilde{T} \rangle$ for each value (stage) of $F$ considered in simulations $\mathcal{S}_{1}^1$ and $\mathcal{S}_{3}^1$. When $F<F_c$, i.e., $F$ is subcritical, there is no motion nor mixing such that the dimensionless temperature profile is fully conductive, i.e., $\widetilde{T}=1-z/h$, as shown by the black solid lines. As $F>F_c$ is increased (dark to light colors; following the direction of the arrow), convective motions emerge, intensify and mix the lake's unstable bulk more and more efficiently. The increased mixing results in a decreasing temperature of the lake's bulk and a decreasing temperature of the bottom boundary. For simulation $\mathcal{S}_1^1$ (figure \ref{fig3}(c)), the temperature profile remains conductive, i.e., linearly decreasing, in the top stably-stratified layer where convective motions are inhibited (because $\widetilde{\beta}<0$). The well-mixed convective region is small compared to the stably stratified region in $\mathcal{S}_1^1$ initially, but the situation reverses as $F$ increases. For simulation $\mathcal{S}_3^1$ (figure \ref{fig3}(d)), convection occurs everywhere such that $\langle \widetilde{T} \rangle$ has a top-down symmetry and the bulk temperature is approximately the average of the top and bottom temperatures. We display the time history of the horizontally-averaged bottom temperature $\langle \widetilde{T}_b \rangle_h = \langle \widetilde{T}(z=0) \rangle_h$ of simulations $\mathcal{S}_1^1$ and $\mathcal{S}_3^1$ in figures \ref{fig3}(e) and \ref{fig3}(f), respectively. The bottom temperature decreases in smooth steps every time the heat flux (or, alternatively, the water depth for the 2$^{nd}$ experiment) increases. There are 10 stages in simulation $\mathcal{S}_1^1$, each lasting one diffusive time, and 7 stages in simulation $\mathcal{S}_3^1$. All simulations with ice overburden pressure $p_i<p_*$ are qualitatively similar to simulation $\mathcal{S}_1^1$, whereas simulations with $p_i\geq p_*$ are qualitatively similar to simulation $\mathcal{S}_3^1$. Note that time averaging of all variables of interest is performed over the second half of each simulation stage.

\subsection{Effective temperature difference and thermal expansion coefficient driving the convection}\label{sec:eff} 
 
{The mean temperature on the bottom boundary is a key output of the simulations since it gives the range of temperatures involved in convective motions and contributing to the heat transport. The effective temperature difference driving the convection, which we denote by $\Delta_{eff}$, may be taken as the difference between the mean bottom temperature and the maximum of the temperature of maximum density and freezing temperature, i.e., in dimensionless form,
\ba{}\label{eq:deltaeff}
\widetilde{\Delta}_{eff}=\Delta_{eff}/\Delta=\langle\widetilde{T}_b\rangle-\widetilde{T}_d(\widetilde{T}_d>0),
\ea
as $\widetilde{\beta}>0$ for $\langle \widetilde{T}_b \rangle\geq\widetilde{T}\geq \langle \widetilde{T}_b \rangle -\widetilde{\Delta}_{eff}$.} Note that the term $(\widetilde{T}_d>0)$ in \eqref{eq:deltaeff} is to be understood as a Heaviside function, i.e., such that it is 1 if $\widetilde{T}_d>0$ and 0 otherwise (in fact, all greater than or less than signs in between parentheses should be understood as Heaviside functions in this paper). For simulations with $\widetilde{T}_d<0$, i.e., which are fully convective, the dimensionless effective temperature difference is simply equal to the dimensionless mean bottom temperature. For simulations with $\widetilde{T}_d>0$, there can be no convection in the temperature range $0<\widetilde{T}<\widetilde{T}_d$, such that the effective temperature difference is equal to the mean bottom temperature minus the temperature of maximum density. {Since the mean bottom temperature is the highest (on average) temperature in the lake, it sets not only the effective temperature difference driving the convection but also the maximum value of the thermal expansion coefficient, which we write in dimensionless form as $\widetilde{\beta}_b=\widetilde{\beta}(\langle\widetilde{T}_b\rangle)={\beta}_b/\overline{\beta}_b$ with subscript $_b$ denoting bottom variables.} We recall that $\widetilde{\beta}$ is also a function of $p_i$. However, the ice overburden pressure is fixed for each simulation, such that its influence on $\widetilde{\beta}$ is not shown for simplicity. {The effective thermal expansion coefficient $\widetilde{\beta}_{eff}$ can be taken as the average between the bottom (maximum) thermal expansion coefficient and the thermal expansion coefficient at the top of the convective layer, which is $\widetilde{\beta}_{f}=\widetilde{\beta}(\widetilde{T}_f=0)>0$ if the lake is fully convective and 0 if the lake has a stable layer (since in this case the mean temperature at the top of the convective layer is $\widetilde{T}_d$), viz.
\ba{}\label{eq:betaeff}
\widetilde{\beta}_{eff} = \f{\widetilde{\beta}_b+\widetilde{\beta}_f(\widetilde{T}_d<0)}{2}. 
\ea
}

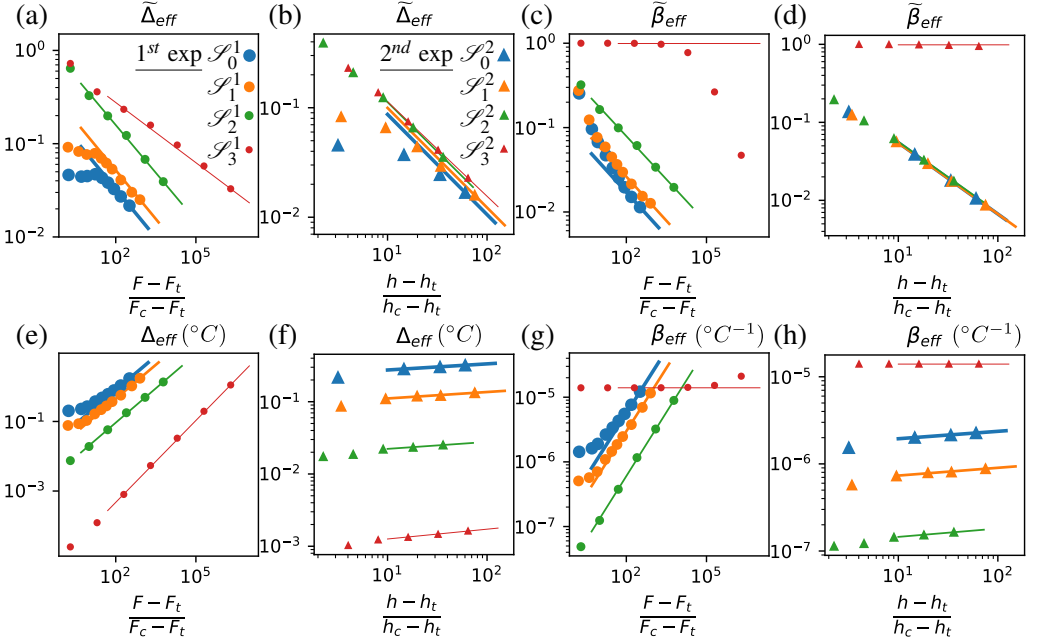
\begin{figure}
\centering
\includegraphics[width=1\textwidth]{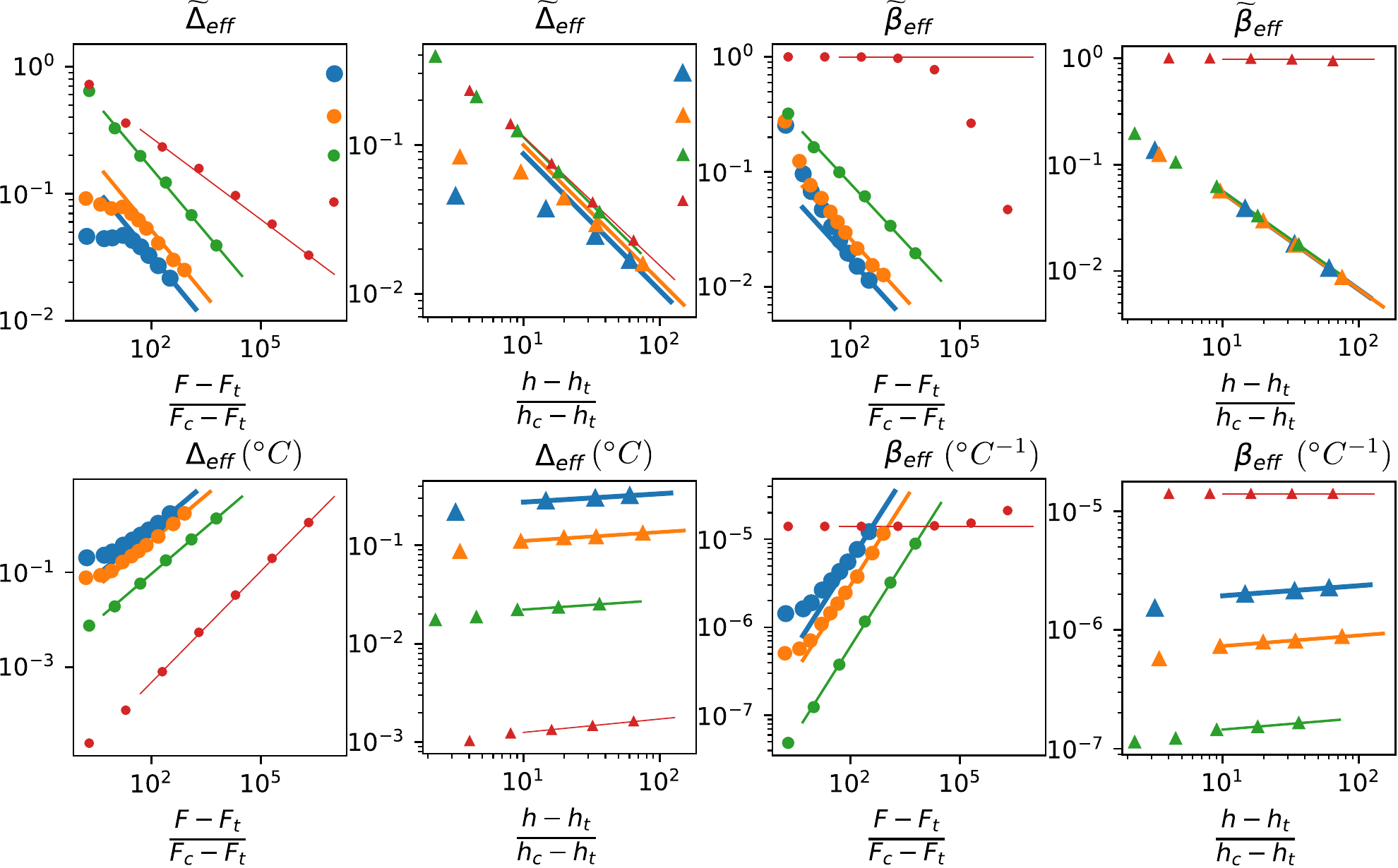}
\put(-335,216){\small{\underline{1$^{st}$ exp}}}
\put(-308,216){\small{$\mathcal{S}_0^1$}}
\put(-308,204){\small{$\mathcal{S}_1^1$}}
\put(-308,192){\small{$\mathcal{S}_2^1$}}
\put(-308,180){\small{$\mathcal{S}_3^1$}}	
\put(-243,216){\small{\underline{2$^{nd}$ exp}}}
\put(-213,216){\small{$\mathcal{S}_0^2$}}
\put(-213,204){\small{$\mathcal{S}_1^2$}}
\put(-213,192){\small{$\mathcal{S}_2^2$}}
\put(-213,180){\small{$\mathcal{S}_3^2$}}
\put(-380,230){\large{(a)}}
\put(-285,230){\large{(b)}}
\put(-190,230){\large{(c)}}
\put(-95,230){\large{(d)}}
\put(-380,110){\large{(e)}}
\put(-285,110){\large{(f)}}
\put(-190,110){\large{(g)}}
\put(-95,110){\large{(h)}}
\vspace{-0.in}\caption{(a),(b) Dimensionless effective temperature difference $\widetilde{\Delta}_{eff}$ (i.e., driving the convection) as a function of (a) the normalized geothermal flux $(F-F_t)/(F_c-F_t)$ for the simulations of the first experiment and (b) the normalized water depth $(h-h_t)/(h_c-h_t)$ for the simulations of the second experiment (cf. legends and table \ref{tab:sims}). (c),(d) same as (a),(b) but for the dimensionless effective thermal expansion coefficient $\widetilde{\beta}_{eff}$. (e)-(h) show the same variables as (a)-(d) but in dimensional form, i.e., with $\Delta_{eff}=\Delta\widetilde{\Delta}_{eff}$ and ${\beta}_{eff}=\overline{\beta}_b\widetilde{\beta}_{eff}$. {The solid lines show scaling laws as discussed in section \S\ref{sec:scalings} and listed in table \ref{tab:scalings}.} {Note that the symbols' size and lines' thickness is inversely proportional to $p_i$, i.e., large (resp. small) symbols and thick (resp. thin) lines highlight results for small (resp. large) $p_i$.}}
\label{fig4}
\end{figure}

We show in figures \ref{fig4}(a)-(d) the evolutions of the dimensionless effective temperature difference $\widetilde{\Delta}_{eff}$ and thermal expansion coefficient $\widetilde{\beta}_{eff}$ as we increase the heat flux $F$ (for the first experiment) or the water depth $h$ (for the second experiment) in the simulations. Figure \ref{fig4}(a) shows that $\widetilde{\Delta}_{eff}$ decreases monotonically with the normalized heat flux $(F-F_t)/(F_c-F_t)$ in all simulations (of the first experiment). Two asymptotic behaviours, highlighted by the solid lines, emerge at relatively large values of $(F-F_t)/(F_c-F_t)$. The asymptotic behaviour is the same for simulations $\mathcal{S}_i^1$ ($i=0,1,2$) but is different for simulation $\mathcal{S}_3^1$. A similar result is obtained  with the second experiment, as shown in figure \ref{fig4}(b), i.e., $\widetilde{\Delta}_{eff}$ decreases monotonically with the normalized water depth $(h-h_t)/(h_c-h_t)$ and display two asymptotic behaviours, although the difference between the asymptotic behaviour for $\mathcal{S}_i^2$ ($i=0,1,2$) and $\mathcal{S}_3^2$ is tenuous (which is expected, as we will demonstrate in section \S\ref{sec:scalings}). The origin of the two different asymptotic behaviours for $\widetilde{\Delta}_{eff}$ can be related to the evolution of $\widetilde{\beta}_{eff}$  with the normalized heat flux and water depth shown in figure \ref{fig4}(c) and \ref{fig4}(d), respectively. On the one hand, the effective thermal expansion coefficient $\widetilde{\beta}_{eff}$ decreases monotonically and displays a common asymptotic behaviour with the normalized heat flux for simulations $\mathcal{S}_i^1$ with $i=0,1,2$ (figure \ref{fig4}(c)) and with the normalized water depth  for simulations $\mathcal{S}_i^2$ with $i=0,1,2$ (figure \ref{fig4}(d)). On the other hand, $\widetilde{\beta}_{eff}\approx 1$ for simulations $\mathcal{S}_3^1$ (figure \ref{fig4}(c)) and  $\mathcal{S}_3^2$ (figure \ref{fig4}(d)), although it can be seen that $\widetilde{\beta}_{eff}$ starts decreasing with the normalized {heat flux} at large values for simulation $\mathcal{S}_3^1$ (figure \ref{fig4}(c)).

In order to understand why $\widetilde{\beta}_{eff}$ either stagnates or decreases with the normalized heat flux or water depth, it is useful to look at the dimensional effective temperature difference $\Delta_{eff}=\Delta\widetilde{\Delta}_{eff}$ and the dimensional effective thermal expansion coefficient $\beta_{eff}=\overline{\beta}_{b}\widetilde{\beta}_{eff}$ shown in figures \ref{fig4}(e)-(h). Figures \ref{fig4}(e) and \ref{fig4}(f) show that $\Delta_{eff}$ increases in all simulations. This happens because $\widetilde{\Delta}_{eff}$ decreases more slowly with increasing $F$ or $h$ (which increase mixing), than the temperature scale $\Delta=Fh/k$ increases with $F$ or $h$. Since $\Delta_{eff}$ increases in all simulations, the bottom temperature also increases, and so does the thermal expansion coefficient $\beta_{eff}$ (figures \ref{fig4}(g) and \ref{fig4}(h)). However, {$\beta_{eff}=\overline{\beta}_b[\Delta_{eff}+2(T_f-T_d)(T_d<T_f)]/[2(\overline{T}_b-T_d)]$} is an affine function of $\Delta_{eff}$ (cf. equation \eqref{eq:betaeff}). Thus, while we might expect that the effective thermal expansion coefficient always scales asymptotically like $\beta_{eff}\sim \Delta_{eff}$ as $\Delta_{eff} \rightarrow \infty$, there exists a range of effective temperature differences, or bottom temperatures, i.e., $0<\langle T_b\rangle-T_f\ll T_f-T_d$, when ${T}_d<T_f$ for which $\beta_{eff}$ increases negligibly and remains approximately constant. The range of bottom temperatures for which $\beta_{eff}$ can be considered constant shrinks to 0 for subglacial lakes with $p_i\leq p_*$ since ${T}_d\geq T_f$ in this case, and it increases as $p_i>p_*$ increases. The next section (\S\ref{sec:scalings}) presents a derivation of two sets of closed-form expressions for the variables of interest, including the bottom temperature, in terms of the problem parameters, based on whether we assume that $\beta_{eff}$ is constant or is linearly proportional to $\Delta_{eff}$. Then, section \S\ref{sec:discussion}  discusses which predictions are applicable to subglacial lakes in Antarctica.

\subsection{Predictions in the limit of decoupled convective and stably stratified dynamics}\label{sec:scalings}

In this section we set out to define an effective (output) Rayleigh number $Ra_{eff}$ based on the simulation results and explore its dependence with the control parameters. We then demonstrate that the Nusselt number, $Nu$, which estimates the contribution of fluid motions to the transport of heat relative to conduction alone, and the Reynolds number, $Re$, which compares fluid inertia to viscous dissipation, display asymptotic behaviours with $Ra_{eff}$ similar to those observed in classical Rayleigh-B\'enard convection. This allows the derivation of predictive expressions for all output variables of interest in terms of the control parameters.

{We assume that the convective and stably-stratified layers are dynamically decoupled}, such that we can simply define the effective Rayleigh number  as
\ba{} \label{eq:Raeff}
Ra_{eff} = \f{g\Delta_{eff}\beta_{eff} h_{eff}^3}{\nu\kappa},
\ea
where $h_{eff}$ is the effective (decoupled) convective layer depth and $\Delta_{eff}$ and $\beta_{eff}$ are the effective temperature difference and thermal expansion coefficient discussed in section \S\ref{sec:eff} and whose dimensionless forms are given by equations \eqref{eq:deltaeff} and \eqref{eq:betaeff}. {Neglecting the influence from the stable layer on the convection means that the effective convective layer depth is well approximated by the full depth minus the mean thickness of the top stably-stratified layer, which is equal to $h(T_d-T_f)/\Delta$ if $T_d>T_f$ and 0 otherwise.} Thus, 
\ba{}\label{eq:heff}
h_{eff} = h\lb 1 -\f{T_d-T_f}{\Delta}\lp T_d>T_f \rp\rb,
\ea
which can be rewritten as{
\ba{}\label{eq:heff2}
h_{eff} = h\lb 1 -\widetilde{T}_d(\widetilde{T}_d>0) \rb = \f{h}{1+\overline{S}(\overline{S}>0)},
\ea
}using either the dimensionless temperature of maximum density or the base-state stiffness parameter (cf. equation \eqref{eq:cpar}). {Using \eqref{eq:heff2} we can rewrite $Ra_{eff}$, i.e., equation \eqref{eq:Raeff}, as 
\ba{}\label{eq:Raeff2}
Ra_{eff} = \overline{Ra}_F\f{\widetilde{\Delta}_{eff}\widetilde{\beta}_{eff}}{\lb 1+\overline{S}(\overline{S}>0)\rb^3},
\ea
where $\overline{Ra}_F$ is the base-state flux-based Rayleigh number (cf. equation \eqref{eq:cpar}).} Equations \eqref{eq:Raeff} and \eqref{eq:Raeff2} can be expected to  represent accurately the effective Rayleigh number when there is (almost) no contribution from the stable layer to the convective dynamics, but to be inaccurate when the stable layer is entrained into and modifies the properties of the lower convective bulk. \cite{Couston2017} investigated the dynamics of mixed convective and stably-stratified fluids over a broad range of input stiffnesses and showed that the Nusselt number was indeed enhanced due to an entrainment heat flux in the limit of small (effective) stiffness, suggesting that the effective Rayleigh number should be adjusted when entrainment from the stable layer into the convective bulk is significant. {We will show in section \S\ref{sec:entrainment} that the entrainment heat flux is small in all our simulations, such that the assumption of decoupled convective and stably-stratified dynamics, leading to equations \eqref{eq:Raeff} and \eqref{eq:Raeff2} for $Ra_{eff}$, is accurate {at leading order}.} Note that the base-state stiffness $\overline{S}$, which can be related to the effective convective layer depth (cf. \eqref{eq:heff2}), is not the effective stiffness of the system due to the variability of the bottom temperature and thermal expansion coefficient. The effective stiffness of the simulations is estimated \textit{a posteriori} and discussed in section \S\ref{sec:entrainment}.

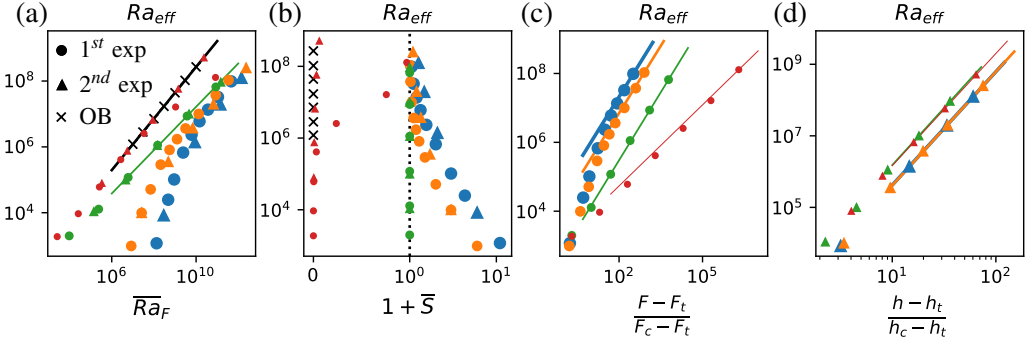
\begin{figure}
\centering
\includegraphics[width=1\textwidth]{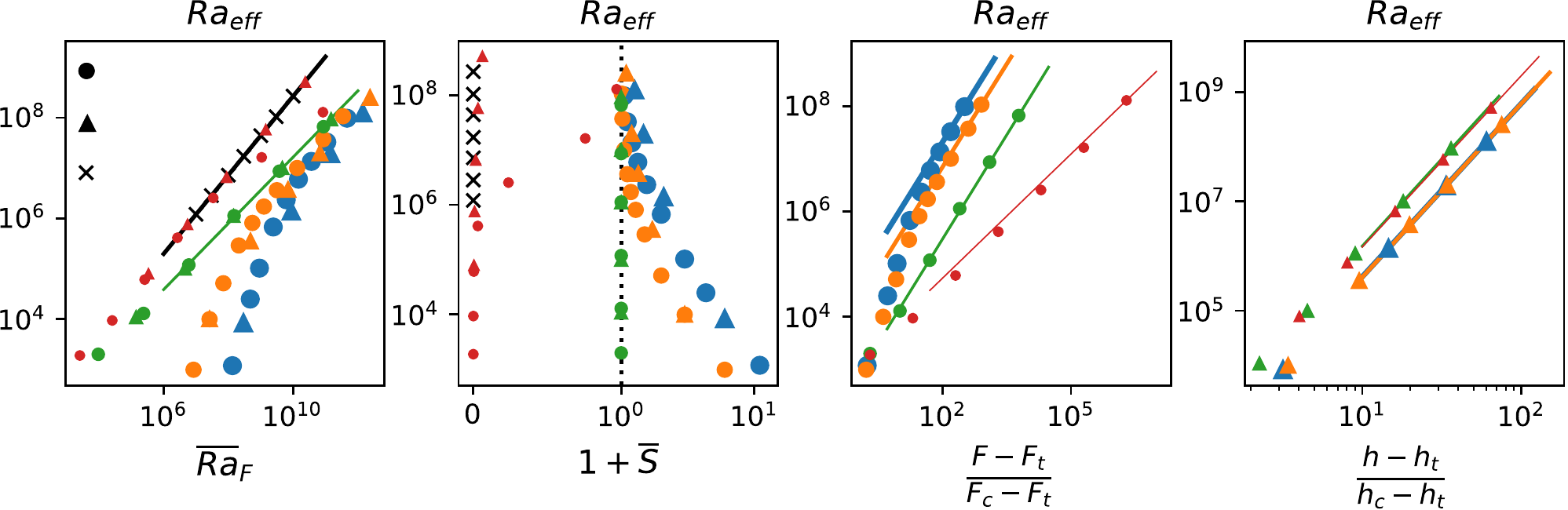}
\put(-356,105){\small{{1$^{st}$ exp}}}
\put(-356,92){\small{{2$^{nd}$ exp}}}
\put(-356,80){\small{{OB}}}
\put(-380,118){\large{(a)}}
\put(-285,118){\large{(b)}}
\put(-190,118){\large{(c)}}
\put(-95,118){\large{(d)}}
\vspace{-0.05in}\caption{(a) Effective Rayleigh number $Ra_{eff}$ as a function of $\overline{Ra}_F$ for all simulations, including the Oberbeck-Boussinesq (OB) simulations $\mathcal{S}_{OB}$ whose results are shown by black crosses. Note that, as in figure \ref{fig4}, large (resp. small) symbols highlight simulation results with small (resp. large) $p_i$. {The thick black and thin green solid lines show the predictive expressions for $Ra_{eff}$ as a function of $\overline{Ra}_F$ for simulations $\mathcal{S}_{OB}$ (OB regime) and $\mathcal{S}_{2}^1$ (LEC regime), respectively (cf. tables \ref{tab:coeffs} and \ref{tab:scalings})}. (b) $Ra_{eff}$ as a function of $1+\overline{S}$. The vertical dotted line highlights $\overline{S}=0$. (c) $Ra_{eff}$ for the first experiment only as a function of the normalized heat flux $(F-F_t)/(F_c-F_t)$. (d) $Ra_{eff}$ for the second experiment only as a function of the normalized water depth $(h-h_t)/(h_c-h_t)$. {The solid lines in (c) and (d) show the theoretical scalings for $Ra_{eff}$ with $F$ and $h$ (cf. table \ref{tab:scalings}) in the OB regime (thin red lines) and in the LEC regime (all other lines) with $\alpha$ and $\gamma$ listed in table \ref{tab:coeffs}.}}
\label{fig5}
\end{figure}

Figures \ref{fig5}(a)-(d) show the effective Rayleigh number $Ra_{eff}$ as a function of the control parameters $\overline{Ra}_F$ and $\overline{S}$ or normalized heat flux $(F-F_t)/(F_c-F_t)$ and water depth $(h-h_t)/(h_c-h_t)$. Figures \ref{fig5}(a) and \ref{fig5}(b) show $Ra_{eff}$ for all simulations, i.e., combining the results of the first and second experiments, {in addition to the effective Rayleigh number for a series of Oberbeck-Boussinesq (OB) simulations, which we denote by $\mathcal{S}_{OB}$ and which have different $\overline{Ra}_{F}$ and fixed $Pr=12.8$ (shown by the black crosses).} {The definition of the effective Rayleigh number for $\mathcal{S}_{OB}$ is simply $Ra_{eff}=\widetilde{\Delta}_{eff}\overline{Ra}_F$ since $\overline{S}=0$ and all physical variables, including the thermal expansion coefficient, are assumed constants in the OB approximation.} It can be seen in figure \ref{fig5}(a) that $Ra_{eff}$ in simulations $\mathcal{S}_3^2$ (red triangles) and $\mathcal{S}_3^1$ (red circles; for relatively low values of $\overline{Ra}_F$) follows the same trend as $Ra_{eff}$ in simulation $\mathcal{S}_{OB}$ (black crosses), whose asymptotic behaviour is shown by the black solid line. This suggests that subglacial lakes that are fully convective, and for which the effective thermal expansion coefficient is independent of $\overline{Ra}_{F}$, i.e., constant, behave similarly to OB fluids. For simulations $\mathcal{S}_i^j$ ($i=0,1,2$; $j=1,2$), $Ra_{eff}$ also increases with $\overline{Ra}_{F}$ (whose asymptotic behaviour is shown by the green solid line) but following a trend that is markedly different from the OB results, which is due to the variability of the effective thermal expansion coefficient with the control parameters when $\overline{S}\geq 0$ (or $p_i\leq p_*$). Figure \ref{fig5}(b) shows $Ra_{eff}$ as a function of $(1+\overline{S})$ instead of $\overline{S}$ because the base-state stiffness enters the definition of $Ra_{eff}$ as $1/(1+\overline{S})$ (cf. equation \eqref{eq:heff2}). It should be noted that  $\overline{Ra}_{F}$ and $\overline{S}$ are not independent variables in our two numerical experiments, since we decided to explore the dynamics of subglacial lakes by sweeping along lines of constant heat flux and water depth rather than lines of constant $\overline{Ra}_{F}$ and $\overline{S}$. Thus, the increase of $Ra_{eff}$ for simulation $\mathcal{S}_3^1$ with $\overline{S}$ (small red circles in figure \ref{fig5}(b)) is due to the associated increase of $\overline{Ra}_{F}$, not $\overline{S}$. Similarly, the decrease of $Ra_{eff}$ with increasing $\overline{S}$ for simulations $\mathcal{S}_i^j$ ($i=0,1$; $j=1,2$) is due to the associated decrease of $\overline{Ra}_{F}$ (cf. figure \ref{fig2}). The existence of two distinct scaling behaviours for $Ra_{eff}$ with the problem parameters is again clear in figure \ref{fig5}(c) where $Ra_{eff}$ follows a common asymptotic behaviour for simulations $\mathcal{S}_i^1$ ($i=0,1,2$), shown by the parallel blue, orange and green (relatively thick) solid lines, and one other for simulation $\mathcal{S}_3^1$, shown by the thin red solid line. The difference in asymptotic behaviours for $Ra_{eff}$ between simulations $\mathcal{S}_i^2$ ($i=0,1,2$) and simulation $\mathcal{S}_3^2$ in figure \ref{fig5}(d) is tenuous. This is because the difference in scalings for $Ra_{eff}$ with the water depth is much weaker than with the heat flux, as we demonstrate at the end of this section.

Now that we have explored the dependence of the effective Rayleigh number with the problem parameters, we turn our attention to the Nusselt number $Nu$ and the Reynolds number $Re$. {We define the Nusselt number as the ratio of the full heat flux $F$ divided by the conductive heat flux based on the output temperature difference between the top and bottom boundaries of the convective bulk at statistical steady state, i.e.
\ba{}\label{eq:Nu}
Nu = \f{F}{k\f{ \Delta_{eff}}{ h_{eff}}} = \f{\widetilde{h}_{eff}}{\widetilde{\Delta}_{eff}} = \f{1}{\widetilde{\Delta}_{eff}\lb 1+\overline{S}(\overline{S}>0)\rb},
\ea
with $\widetilde{h}_{eff}=h_{eff}/h$ and the Reynolds number as 
\ba{}\label{eq:Re}
Re = \f{V_{rms}h_{eff}}{\nu},
\ea
with $V_{rms}=\sqrt{h_{eff}^{-1}\int_0^{h_{eff}} \langle |\bold{u}|^2 \rangle dz}$ the root-mean-square (rms) velocity within the convective layer.}

\begin{figure}
\centering
\includegraphics[width=1\textwidth]{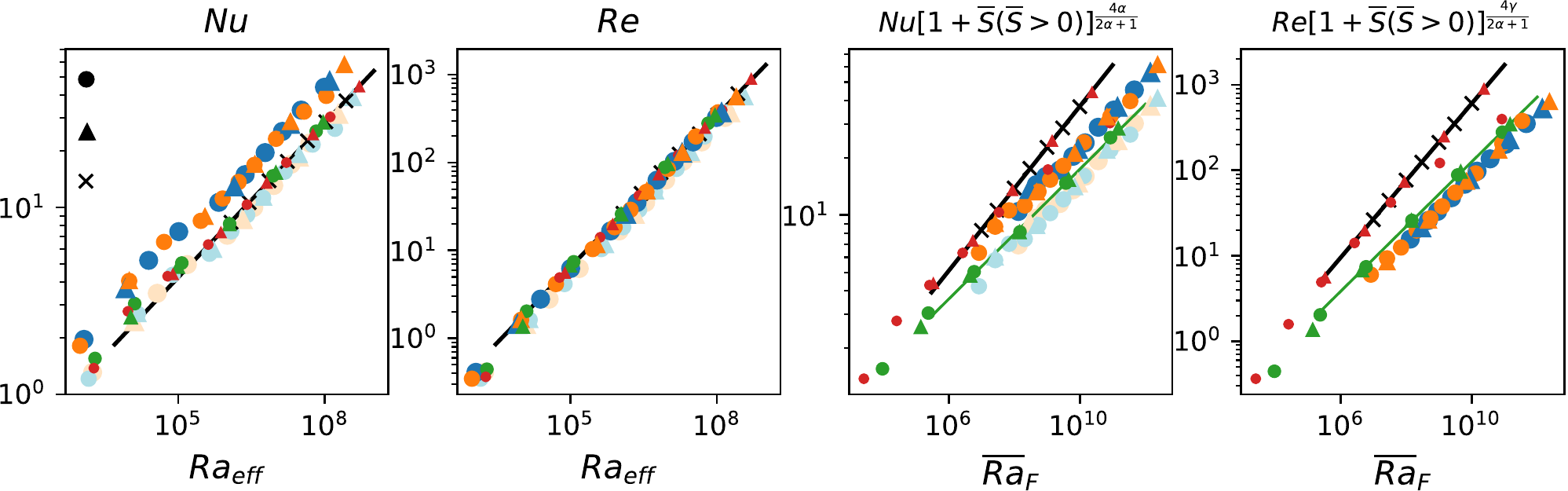}
\put(-355,99){\small{{1$^{st}$ exp}}}
\put(-355,86){\small{{2$^{nd}$ exp}}}
\put(-355,74){\small{{OB}}}
\put(-380,112){\large{(a)}}
\put(-285,112){\large{(b)}}
\put(-190,112){\large{(c)}}
\put(-95,112){\large{(d)}}
\vspace{-0.05in}\caption{(a) Nussel number $Nu$ as a function of $Ra_{eff}$ for all simulations (we use the same symbol colors and size chart as in figures \ref{fig4}-\ref{fig5}). (b) Reynolds number $Re$ as a function of $Ra_{eff}$. {The black solid line in (a) and (b) shows the power law fit \eqref{eq:NuPL} and \eqref{eq:RePL} for $Nu$ and $Re$ for the results of simulation $\mathcal{S}_{OB}$ (cf. table \ref{tab:coeffs}).} (c) Compensated Nusselt number as a function of $\overline{Ra}_F$. (d) Compensated Reynolds number  as a function of $\overline{Ra}_F$. {The black (resp. green) solid lines in (c) and (d) show the power law fits provided in table \ref{tab:scalings} for the results of simulation $\mathcal{S}_{OB}$ (resp. $\mathcal{S}_2^1$), i.e., in the OB regime (resp. LEC regime).} The light-colored markers in (a), (b) and (c) show the results for simulations $\mathcal{S}_{i}^{j}$ ($i=0,1$; $j=1,2$) with $Nu$ multiplied by 2/3 and $Ra_{eff}$ multiplied by 3/2 (see details in section \S\ref{sec:entrainment}).}
\label{fig6}
\end{figure}
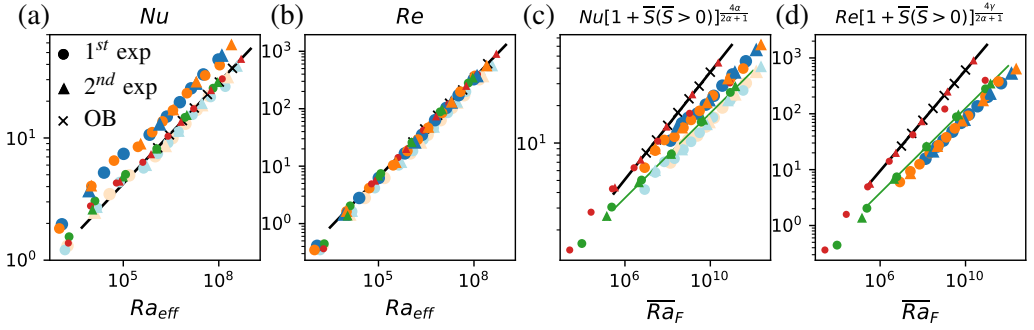

We show the Nusselt number $Nu$ as a function of $Ra_{eff}$ in figure \ref{fig6}(a) for all simulations of the first experiment (with results shown by circles), all simulations of the second experiment (shown by triangles) and for simulation $\mathcal{S}_{OB}$ (black crosses). There is a clear universal asymptotic scaling of $Nu$ with $Ra_{eff}$ in all cases, which is highlighted by the black solid line that shows the best-fit power law for the OB results. More precisely, the Nusselt number in all simulations can be expressed to a good approximation as a power law of the form
\ba{}\label{eq:NuPL}
Nu = a Ra_{eff}^{\alpha},
\ea
where {pre factor $a$ and exponent $\alpha$ are reported in table \ref{tab:coeffs} for each simulation along with the relative error, which is typically of order 1\% and always less than 8\%}. The exponent $0.26<\alpha<0.29$ shows little variability across the simulations, as can be seen from the similar slopes of the simulation results for large $Ra_{eff}$ (cf. figure \ref{fig6}(a)). The pre factor is slightly more variable, i.e., $0.16<a<0.31$, and is relatively large for simulations $\mathcal{S}_{i}^j$ ($i=0,1$; $j=1,2$), i.e., with a stable layer, as can be seen from the upward shift of the large blue and orange circles and triangles relative to the other smaller symbols and the solid black line. {The discrepancy of $Nu$ between simulations that are fully convective and simulations with a top stably-stratified layer is due to the fact that the effective temperature difference $\widetilde{\Delta}_{eff}$ as defined by \eqref{eq:deltaeff} underestimates the range of temperatures involved in the convective heat transport, which is extended in the presence of a stable layer. Substituting $\widetilde{\Delta}_{eff}$ with $3/2\widetilde{\Delta}_{eff}$ in the definitions of $Ra_{eff}$ and $Nu$ helps correct the discrepancy and yields the results shown by the light-blue and light-orange markers in figure \ref{fig6}(a), which better overlap with the other simulation results (cf. details in section \S\ref{sec:entrainment}).} Similar to the Nusselt number, the Reynolds number also follows an almost universal scaling with $Ra_{eff}$ (see figure \ref{fig6}(b)), such that it can be predicted to a good approximation using a power law of the form
\ba{}\label{eq:RePL}
Re = c Ra_{eff}^{\gamma},
\ea
with almost the same pre factor $c$ and exponent $\gamma$ across all simulations (cf. table \ref{tab:coeffs}). {Note that the standard deviation $\sigma_{Re}$ (in time) of the Reynolds number (resp.  $\sigma_{Nu}$ for the Nusselt number) over the (second) half of the diffusive time scale of each simulation stage is always in the range $10^{-2}Re<\sigma_{Re}<10^{-1}Re$ (resp. $10^{-3}Nu<\sigma_{Nu}<3\times 10^{-2}Nu$), i.e., small, except for simulation $\mathcal{S}_3^1$ with the smallest heat flux, for which $\sigma_{Re}\approx 1.4Re$, because $Re$ increases slowly toward its final value $Re\approx 0.5$ in this case.}

\begin{table}\centering 
\begin{tabular}{C{1cm}C{1cm}C{1cm}C{1cm}C{1cm}C{1cm}C{1cm}}
s.n. & $a$ & $\alpha$ & rel. err. & $c$ & $\gamma$ & rel. err. \\ 
$\mathcal{S}_0^1$ & 0.31 & 0.27 & 8\%   & 0.0071 & 0.58 & 6\% \\
$\mathcal{S}_1^1$ & 0.30 & 0.27 & 5\%   & 0.0046 & 0.61 & 4\% \\
$\mathcal{S}_2^1$ & 0.23 & 0.26 & 6\%   & 0.0088 & 0.58 & 3\% \\
$\mathcal{S}_3^1$ & 0.18 & 0.27 & 0.6\% & 0.0080 & 0.58 & 2\% \\
$\mathcal{S}_0^2$ & 0.22 & 0.29 & 0.6\% & 0.0060 & 0.59 & 3\% \\
$\mathcal{S}_1^2$ & 0.23 & 0.29 & 2\%   & 0.0062 & 0.59 & 3\% \\
$\mathcal{S}_2^2$ & 0.16 & 0.28 & 0.7\% & 0.0075 & 0.59 & 2\% \\
$\mathcal{S}_3^2$ & 0.18 & 0.28 & 0.9\% & 0.0075 & 0.58 & 3\% \\
$\mathcal{S}_{OB}$ & 0.18 & 0.28 & 1\% & 0.0086 & 0.58 & 2\% \\
\hline
\end{tabular}\vspace{-0.in}\caption{Best-fit coefficients for the power laws $Nu=a Ra_{eff}^{\alpha}$ and $Re=cRa_{eff}^{\gamma}$ for $Ra_{eff}>10^5$ for all simulations, including the Oberbeck-Boussinesq simulation $\mathcal{S}_{OB}$. s.n. means simulation name while rel. err. denotes the maximum relative error between the simulation results and the predictive best-fit power law.}\label{tab:coeffs}\end{table}

With the predictive power law \eqref{eq:NuPL} for $Nu$ in terms of $Ra_{eff}$ in hand, we can derive a predictive equation for the effective temperature difference $\widetilde{\Delta}_{eff}$ in terms of the problem parameters. Substituting the expressions \eqref{eq:Raeff2} and \eqref{eq:Nu} for $Ra_{eff}$ and $Nu$ into \eqref{eq:NuPL}, we obtain
\ba{}\label{eq:der1}
\f{1}{\widetilde{\Delta}_{eff}\lb 1+\overline{S}(\overline{S}>0)\rb } = a \lcb \overline{Ra}_F\f{\widetilde{\Delta}_{eff}\widetilde{\beta}_{eff}}{\lb 1+\overline{S}\lp\overline{S}>0\rp\rb^3} \rcb^{\alpha},
\ea
which is an equation between $\widetilde{\Delta}_{eff}$, $\widetilde{\beta}_{eff}$ and the control parameters. The effective thermal expansion coefficient \eqref{eq:betaeff} can be expressed in terms of $\widetilde{\Delta}_{eff}$ and $\overline{S}$ as 
\ba{}\label{eq:der2}
\widetilde{\beta}_{eff} = \f{\widetilde{\Delta}_{eff}}{2}\lp 1+\overline{S} \rp  - \overline{S}\lp \widetilde{T}_d < 0 \rp,
\ea
with the second term on the right-hand side being the non-zero dimensionless thermal thermal expansion coefficient at the top of subglacial lakes that are fully convective, i.e., for which $\widetilde{T}_d<0$ or $-1\leq \overline{S}< 0$. Substituting \eqref{eq:der2} into \eqref{eq:der1} yields an algebraic equation for $\widetilde{\Delta}_{eff}$ in terms of the control parameters $\overline{Ra}_F$ and $\overline{S}$, viz.{
\ba{}\label{eq:der3}
\widetilde{\Delta}_{eff}^{1+\alpha} \lb \f{\widetilde{\Delta}_{eff}}{2}\lp 1+\overline{S} \rp - \overline{S}\lp  \widetilde{T}_d < 0 \rp \rb^{\alpha} =  \f{\lb 1+\overline{S}\lp\overline{S}>0\rp\rb^{3\alpha-1}}{a \overline{Ra}_F^{\alpha}}.
\ea
}Equation \eqref{eq:der3} is valid as long as the power law \eqref{eq:NuPL} provides a good approximation to the simulation results but is nonlinear in $\widetilde{\Delta}_{eff}$. {Based on the results of section \S\ref{sec:eff}, we know that the thermal expansion coefficient in subglacial lakes exhibits two limiting behaviours, i.e., $\beta_{eff}$ is approximately constant, which we refer to as the Oberbeck-Boussinesq (OB) regime, or $\beta_{eff}$ is linearly proportional to the effective temperature difference, which we refer to as the {linear} expansion coefficient (LEC) regime.} Equation \eqref{eq:der3} can be simplified in both regimes. In the OB regime, $\beta_{eff}=\overline{\beta}_b$, i.e., $\widetilde{\beta}_{eff}=1$, and $-1\leq\overline{S}<0$ (or $\widetilde{T}_d<0$), such that equation \eqref{eq:der3} reduces to {
\ba{}\label{eq:der4}
\widetilde{\Delta}_{eff}^{1+\alpha} = \f{1}{a\overline{Ra}_F^{\alpha}}. 
\ea
}In the LEC regime, $\overline{S}>0$ ($\widetilde{T}_d>0$) or $|\overline{S}|\ll \widetilde{\Delta}_{eff}\leq 1$, such that the second term in equation \eqref{eq:der2} is negligible and equation \eqref{eq:der3} can be approximated as {
\ba{}\label{eq:der5}
\widetilde{\Delta}_{eff}^{1+2\alpha} = \f{2^{\alpha}\lp 1+\overline{S} \rp^{2\alpha-1}}{a\overline{Ra}_F^{\alpha}}. 
\ea
}Substituting the expressions \eqref{eq:cpar} for $\overline{Ra}_F$ and $\overline{S}$ in terms of the control parameters into equation \eqref{eq:der4} yields an asymptotic scaling in the OB regime, which assumes constant $\overline{\beta}_b$ (such that $\overline{Ra}_F\sim Fh^4$), for the dimensional effective temperature difference as
\ba{}\label{eq:der6}
\Delta_{eff} = \Delta \widetilde{\Delta}_{eff} \sim F^{\f{1}{1+\alpha}} h^{\f{1-3\alpha}{1+\alpha}},
\ea
which we recall is related to the mean bottom temperature through $\langle T_b \rangle = T_f + \Delta_{eff}$. Substituting \eqref{eq:cpar} into \eqref{eq:der5} and using the approximation $\overline{S}\approx 1$ in the limit $(Fh)/k\gg T_d$ (large thermal driving) when $\overline{S}>0$, yields instead
\ba{}\label{eq:der7}
\Delta_{eff} = \Delta \widetilde{\Delta}_{eff} \sim F^{\f{1}{2\alpha+1}}h^{\f{1-3\alpha}{2\alpha+1}},
\ea
in the LEC regime (for which $\overline{Ra}_F\sim F^2h^5$). From the closed-form expressions for $\widetilde{\Delta}_{eff}$ \eqref{eq:der4} and \eqref{eq:der5} it is possible to derive the expressions for all variables of interest, including $Ra_{eff}$, $Nu$ and $Re$ but also dimensional variables such as $\beta_{eff}$ and $V_{rms}$, in terms of the problem parameters. We summarize all key expressions for the variables of interest in terms of $\overline{Ra}_F$ and $\overline{S}$ as well as the asymptotic scaling laws with the water depth and heat flux in table \ref{tab:scalings}. Figures \ref{fig6}(c)-(d) show $Nu$ and $Re$ in terms of $\overline{Ra}_F$ compensated such that the influence of the base-state stiffness predicted in table \ref{tab:scalings} is removed.

{The asymptotic scalings with $F$ and $h$ are valid in the limit $(Fh/k)/|T_d-T_f| \ll 1$, i.e., such that $\overline{S}\approx -1$ for the OB regime and $(Fh/k)/|T_d-T_f| \gg 1$, i.e., such that $\overline{S}\approx 0$, for the LEC regime. The latter is satisfied when $F \rightarrow \infty$ and $h \rightarrow \infty$, i.e., for increasingly large heat flux and water depth. The former limit for the OB regime is only valid for a finite range of heat fluxes and water depth as the thermal expansion coefficient will eventually increase substantially as $F$ and $h$ keeps increasing. The results of simulation $\mathcal{S}_3^1$ show how $\widetilde{\beta}_{eff}$ in figure \ref{fig4}(c) and $Re$ in figure \ref{fig6}(d) transitions from varying according to the OB regime for relatively small $F$ but according to the LEC regime for relatively large $F$.}

The validity of the scaling laws provided in table \ref{tab:scalings} can be verified from the good overlap between the simulation results and the solid lines, which show the asymptotic scalings with $F$ and $h$, in figures \ref{fig4}, \ref{fig5}(c) and \ref{fig5}(d). The relatively thick solid lines that are blue, orange and green, display the asymptotic scalings in the LEC regime, whereas thin red and black solid lines display asymptotic scalings in the OB regime. The simulation results have provided not only the scaling exponents $\alpha$ and $\gamma$, but also the pre factors $a$ and $c$ for the predictive power laws \eqref{eq:NuPL} and \eqref{eq:RePL}. Thus, we can predict the actual value of any variable of interest using the equations listed in table \ref{tab:scalings} rather than just the scalings with the problem parameters. The validity of the full expressions derived for $Ra_{eff}$, $Nu$ and $Re$ can be verified from the good overlap between the simulation results and the black and green solid lines, which show the power law fits, in figure \ref{fig5}(a) and figures \ref{fig6}(a)-(d).

{Here we find $\alpha \approx 2/7$, which was first proposed as a natural scaling exponent for $Nu$ based on phenomenological arguments \cite[][]{Castaing1989},} and $\gamma\approx 3/5$ for all simulations (cf. table \ref{tab:coeffs}). {Substituting $\alpha = 2/7$ and $\gamma = 3/5$ in the asymptotic expressions for the dimensional variables with $F$ and $h$ in table \ref{tab:scalings} yields in the OB regime
\ba{}
h_{eff} \sim h, \quad \Delta_{eff} \sim F^{\f{7}{9}}h^{\f{1}{9}}, \quad \beta_{eff} \sim 1, \quad V_{rms} \sim F^{\f{7}{15}}h^{\f{13}{15}},
\ea
and in the LEC regime
\ba{}
h_{eff} \sim h, \quad \Delta_{eff} \sim F^{\f{7}{11}}h^{\f{1}{11}}, \quad \beta_{eff} \sim F^{\f{7}{11}}h^{\f{1}{11}}, \quad
V_{rms} \sim F^{\f{42}{55}}h^{\f{10}{11}}.
\ea
}It can be seen that the difference in scalings is typically greater with $F$ than with $h$, i.e., for instance, the difference in scaling exponent for $\Delta_{eff}$ with $F$ is $7/9-7/11\approx 0.14$ whereas it is only $1/9-1/11\approx 0.02$ with $h$. The discrepancy in the difference of scaling exponents is the reason the two asymptotic regimes are usually easier to identify in simulation results from the 1st experiment than from the 2nd experiment (compare, e.g., figures \ref{fig4}(a) and (b) and figures \ref{fig5}(c) and (d)). We find that $V_{rms}$ has a  scaling with $F$ and $h$ that is steeper for subglacial lakes in the LEC regime than in the OB regime. The steeper scaling for $V_{rms}$ in the LEC regime occurs because the effective Rayleigh number $Ra_{eff}$ increases more rapidly with $F$ and $h$ in the LEC regime, due to the combined increase of the effective thermal driving and thermal expansion coefficient, than in the OB regime for which the thermal expansion coefficient remains constant.

\begin{table}\centering 
\begin{tabular}{C{5cm}C{0.5cm}C{7cm}}
OB regime & & LEC regime \\
\cmidrule{1-1}\cmidrule{3-3}
$\overline{Ra}_F \sim Fh^4$  &  &  $\overline{Ra}_F \sim F^2h^5$ \\ 
$\widetilde{h}_{eff} = 1  \sim 1 $  &  &  $\widetilde{h}_{eff} = \lp 1+\overline{S} \rp^{-1} \sim 1 $  \\ 
$\widetilde{\Delta}_{eff} = a^{\f{-1}{1+\alpha}} \overline{Ra}_F^{\f{-\alpha}{1+\alpha}} \sim \lp Fh^4 \rp^{\f{-\alpha}{1+\alpha}}$  &  &  $\widetilde{\Delta}_{eff} = a^{\f{-1}{2\alpha+1}}2^{\f{\alpha}{2\alpha+1}} \overline{Ra}_F^{\f{-\alpha}{2\alpha+1}} \lp 1+\overline{S} \rp^{\f{2\alpha-1}{2\alpha+1}} \sim \lp F^2h^5 \rp^{\f{-\alpha}{2\alpha+1}}$ \\ 
$\widetilde{\beta}_{eff} = 1 \sim 1 $  &  & $\widetilde{\beta}_{eff} = a^{\f{-1}{2\alpha+1}}2^{\f{-1-\alpha}{2\alpha+1}} \overline{Ra}_F^{\f{-\alpha}{2\alpha+1}} \lp 1+\overline{S} \rp^{\f{4\alpha}{2\alpha+1}} \sim \lp F^2h^5 \rp^{\f{-\alpha}{2\alpha+1}}$  \\ 
$Ra_{eff} = a^{\f{-1}{1+\alpha}} \overline{Ra}_F^{\f{1}{1+\alpha}} \sim \lp Fh^4 \rp^{\f{1}{1+\alpha}}$  &  &  $Ra_{eff} = a^{\f{-2}{2\alpha+1}}2^{\f{-1}{2\alpha+1}} \overline{Ra}_F^{\f{1}{2\alpha+1}} \lp 1+\overline{S} \rp^{\f{-4}{2\alpha+1}} \sim \lp F^2h^5 \rp^{\f{1}{2\alpha+1}}$ \\ 
$Nu = a^{\f{1}{1+\alpha}} \overline{Ra}_F^{\f{\alpha}{1+\alpha}} \sim \lp Fh^4 \rp^{\f{\alpha}{1+\alpha}} $  &  & $Nu = a^{\f{1}{2\alpha+1}}2^{\f{-\alpha}{2\alpha+1}} \overline{Ra}_F^{\f{\alpha}{2\alpha+1}} \lp 1+\overline{S} \rp^{\f{-4\alpha}{2\alpha+1}} \sim \lp F^2h^5 \rp^{\f{\alpha}{2\alpha+1}}$  \\ 
$Re = c a^{\f{-\gamma}{1+\alpha}} \overline{Ra}_F^{\f{\gamma}{1+\alpha}} \sim \lp Fh^4 \rp^{\f{\gamma}{1+\alpha}} $  &  & $Re = c a^{\f{-2\gamma}{2\alpha+1}}2^{\f{-\gamma}{2\alpha+1}} \overline{Ra}_F^{\f{\gamma}{2\alpha+1}} \lp 1+\overline{S} \rp^{\f{-4\gamma}{2\alpha+1}} \sim \lp F^2h^5 \rp^{\f{\gamma}{2\alpha+1}} $  \\ 
$ h_{eff} = h \sim h $  &  & $ h_{eff} = h\lp 1+\overline{S} \rp^{-1} \sim h $ \\ 
$ \Delta_{eff} = \Delta \widetilde{\Delta}_{eff} \sim F^{\f{1}{1+\alpha}}h^{\f{1-3\alpha}{1+\alpha}} $  &  & $ \Delta_{eff} = \Delta \widetilde{\Delta}_{eff} \sim F^{\f{1}{2\alpha+1}}h^{\f{1-3\alpha}{2\alpha+1}} $  \\ 
$ \beta_{eff} = \overline{\beta}_b \widetilde{\beta}_{eff}  \sim 1 $  &  &  $ \beta_{eff} = \overline{\beta}_b \widetilde{\beta}_{eff}  \sim F^{\f{1}{2\alpha+1}}h^{\f{1-3\alpha}{2\alpha+1}} $ \\ 
$ V_{rms} \sim F^{\f{\gamma}{1+\alpha}}h^{\f{4\gamma}{1+\alpha}-1} $  &  & $ V_{rms} \sim F^{\f{2\gamma}{2\alpha+1}}h^{\f{5\gamma}{2\alpha+1}-1} $  \\ 
\hline
\end{tabular}\vspace{-0.in}\caption{Key expressions in terms of the control parameters $\overline{Ra}_F$ and $\overline{S}$ (cf. equation \eqref{eq:cpar}) and asymptotic scaling laws in terms of the water depth $h$ and heat flux $F$ for the dimensionless and dimensional variables of interest to the study of turbulent convection in subglacial lakes. The starting point is the algebraic equation for $\widetilde{\Delta}_{eff}$ \eqref{eq:der3}. The difference between the limiting OB regime and LEC regime arises from the independence or variability of the effective thermal expansion coefficient $\widetilde{\beta}_{eff}$ with the effective temperature difference $\widetilde{\Delta}_{eff}$. The asymptotic scalings with $F$ and $h$ are valid in the limit $(Fh/k)/|T_d-T_f| \ll 1$ ($\overline{S}\approx -1$) for the OB regime and $(Fh/k)/|T_d-T_f| \gg 1$ ($\overline{S}\approx 0$) for the LEC regime.}\label{tab:scalings}\end{table}

\subsection{Effective stiffness and entrainment dynamics}\label{sec:entrainment}

The power laws \eqref{eq:NuPL} and \eqref{eq:RePL} for $Nu$ and $Re$ with $Ra_{eff}$ are in excellent agreement with the numerical results using pre factors and exponents that are almost the same across all simulations, including the OB simulation (cf. table \ref{tab:coeffs}). As a result, our definition of $Ra_{eff}$, i.e., equation \eqref{eq:Raeff2}, which assumes decoupled convective and stably-stratified dynamics, is a good proxy for the true effective Rayleigh number of all laboratory-scale subglacial lakes, including lakes that have a top stable layer and can experience penetrative convection. 

In this section, {we investigate in details the influence of the stable fluid layer on the convection in order to predict whether the assumption of decoupled convective and stably-stratified dynamics can hold at significantly higher heat flux and water depth.} {We first estimate the effective stiffness $S_{eff}$ of the convective--stably-stratified interface in the simulations as
\ba{}\label{eq:Seff}
S_{eff} = \f{T_d-T_f}{\langle T_b \rangle-T_d},
\ea
which is the ratio of the temperature difference across the stable layer to the temperature difference across the convective layer. In a mixed convective and stably-stratified subglacial lake, i.e., for which $p_i\leq p_*$, this ratio is equal to the ratio of the opposite of the thermal expansion coefficient at the top of the stable layer (minimum negative thermal expansion coefficient) to the thermal expansion coefficient at the bottom of the convective layer (maximum positive thermal expansion coefficient). Thus, the effective stiffness $S_{eff}$ in \eqref{eq:Seff} is equivalent to the $\Lambda$ parameter in \cite[][]{Toppaladoddi2018} and is similar to the input stiffness parameter $S_i$ in \cite[][]{Couston2017}, although we recall that \cite[][]{Couston2017} used a thermal expansion coefficient that is piecewise constant rather than linearly varying with temperature.} We show $S_{eff}$ as a function of $Ra_{eff}$ in figure \ref{fig7}(a) for all simulations with a stably-stratified layer. The effective stiffness decreases monotonically with $Ra_{eff}$, which is {expected} since the numerator in \eqref{eq:Seff} is constant while the denominator, which is the dimensional thermal driving $\Delta_{eff}$, increases with $Ra_{eff}$ (cf. figures \ref{fig4}(e) and (f)). The decrease of $S_{eff}$ with $Ra_{eff}$ suggests that the stably-stratified layer may modify  the properties of the convective bulk significantly for large enough $F$ and $h$. \cite{Couston2017} demonstrated that the two layers become increasingly coupled as the input stiffness decreases, indeed, with, for instance, the mean temperature of the well-mixed bulk dropping below the temperature of maximum density in their experiment with $Ra=8\times 10^7$ and $S_i=1$.

In order to understand whether the decrease of $S_{eff}$ with $Ra_{eff}$ implies that the stable layer of subglacial lakes modifies the convective bulk dynamics for large (geophysical) $F$ and $h$, as may be expected based on the results of \cite[][]{Couston2017}, we first investigate the evolution of the temperature of the well-mixed bulk, which we define as
\ba{}\label{eq:Tbulk}
\widetilde{T}_{bulk} = \f{1}{\widetilde{h}_{eff}}\int_0^{\widetilde{h}_{eff}} \widetilde{T} d\widetilde{z},
\ea
with $Ra_{eff}$. If there is no stable layer ($\widetilde{T}_d<0$) or if the stable layer does not influence the convective dynamics, we expect $[\widetilde{T}_{bulk}-\widetilde{T}_d(\widetilde{T}_d>0)]\approx [\langle \widetilde{T}_b\rangle-\widetilde{T}_d(\widetilde{T}_d>0)]/2$, i.e., the bulk temperature is the average of the temperatures at the top and bottom of the convection zone. Figure \ref{fig7}(b) shows that the bulk temperature is the average (approximately) of the top and bottom temperatures for simulations without a stable layer (small green and red symbols). For simulations with a stable layer, however, we find that the bulk temperature minus $\widetilde{T}_d(\widetilde{T}_d>0)$ of the two-layer system is typically lower than $[\langle \widetilde{T}_b\rangle-\widetilde{T}_d(\widetilde{T}_d>0)]/2$. Specifically, we find that the bulk temperature decreases initially with $Ra_{eff}$ and then reaches the plateau $[\widetilde{T}_{bulk}-\widetilde{T}_d(\widetilde{T}_d>0)]\approx [\langle \widetilde{T}_b\rangle-\widetilde{T}_d(\widetilde{T}_d>0)]/4$ (black dashed line). For simulation $\mathcal{S}_1^1$, the normalized bulk temperature appears to anomalously increase with $Ra_{eff}$ at large $Ra_{eff}$ (two rightmost orange circles). The increase of the normalized bulk temperature toward the 0.5 mark at large $Ra_{eff}$ for $\mathcal{S}_1^1$ is due to the shrinking of the top stable layer. In fact, $Ra_{eff}$ increases because $F$ increases in the simulations of the first experiment, which forces a thinning of the stratified top layer. As the stably-stratified layer shrinks, possibly to a point where it is thinner than the thermal boundary layer, upward plumes attempting to penetrate into the stable layer rapidly feels the effect of the top wall and lose their inertia, which reduces entrainment of the stable fluid and lowering of the bulk temperature. The vanishing of the stable layer thickness can be seen in figure \ref{fig7}(c), which shows that the convective layer depth $\widetilde{h}_{eff}\rightarrow 1$ as $S_{eff}$ decreases, or, equivalently, as $Ra_{eff}$ increases. The observation of a plateau for $[\widetilde{T}_{bulk}-\widetilde{T}_d(\widetilde{T}_d>0)]/[\langle \widetilde{T}_b\rangle-\widetilde{T}_d(\widetilde{T}_d>0)]$ at 0.25 with $Ra_{eff}$ (let alone possible increase toward the 0.5 mark when the stable layer becomes too thin) suggests that there is an influence of the stable layer on the convection, but that this influence does not increase with increasing $Ra_{eff}$ or decreasing $S_{eff}$. We further demonstrate in the next paragraph that the entrainment of stable fluid into the convection zone does not increase with $Ra_{eff}$, such that the stable layer's influence on the convection is indeed limited and negligible at leading order.

\begin{figure}
\centering
\includegraphics[width=0.8\textwidth]{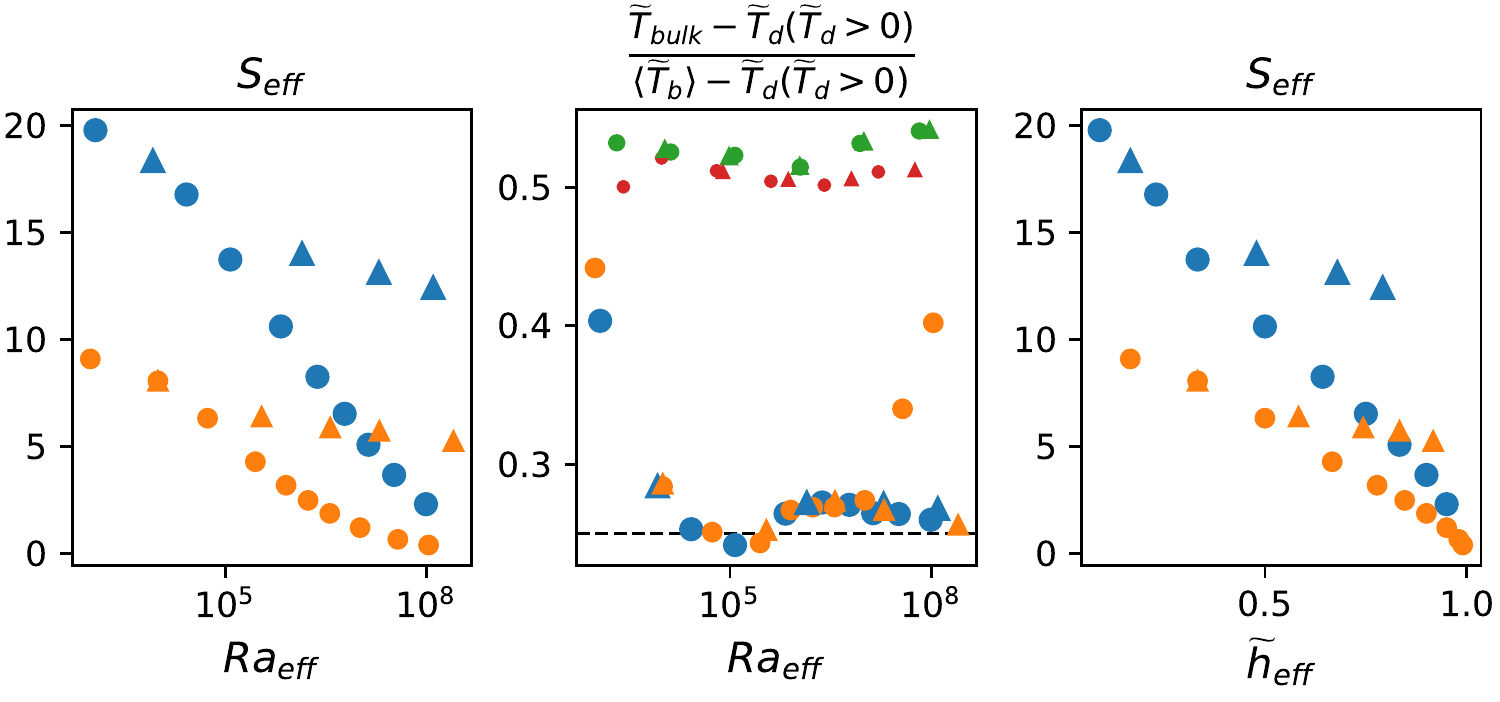}
\put(-310,125){\large{(a)}}
\put(-205,125){\large{(b)}}
\put(-100,125){\large{(c)}}
\vspace{-0.05in}\caption{(a) Effective stiffness $S_{eff}$ (cf. equation \eqref{eq:Seff}) as a function of $Ra_{eff}$ for simulations $\mathcal{S}_i^j$ ($i=0,1$; $j=1,2$), i.e., with a top stably-stratified layer. (b) Ratio between the bulk temperature $\widetilde{T}_{bulk}$ (cf. equation \eqref{eq:Tbulk}) in excess of $\widetilde{T}_d(\widetilde{T}_d>0)$ and the mean bottom temperature $\langle \widetilde{T}_b \rangle$ in excess of $\widetilde{T}_d(\widetilde{T}_d>0)$ as a function of $Ra_{eff}$. The results are shown for all simulations of the first and second experiments and the dashed line highlights the 1/4 ordinate. (c) $S_{eff}$ as a function of the expected dimensionless convective layer depth $\widetilde{h}_{eff}$ for all simulations with a top stably-stratified layer (cf. equation \eqref{eq:heff2}).}
\label{fig7}
\end{figure}
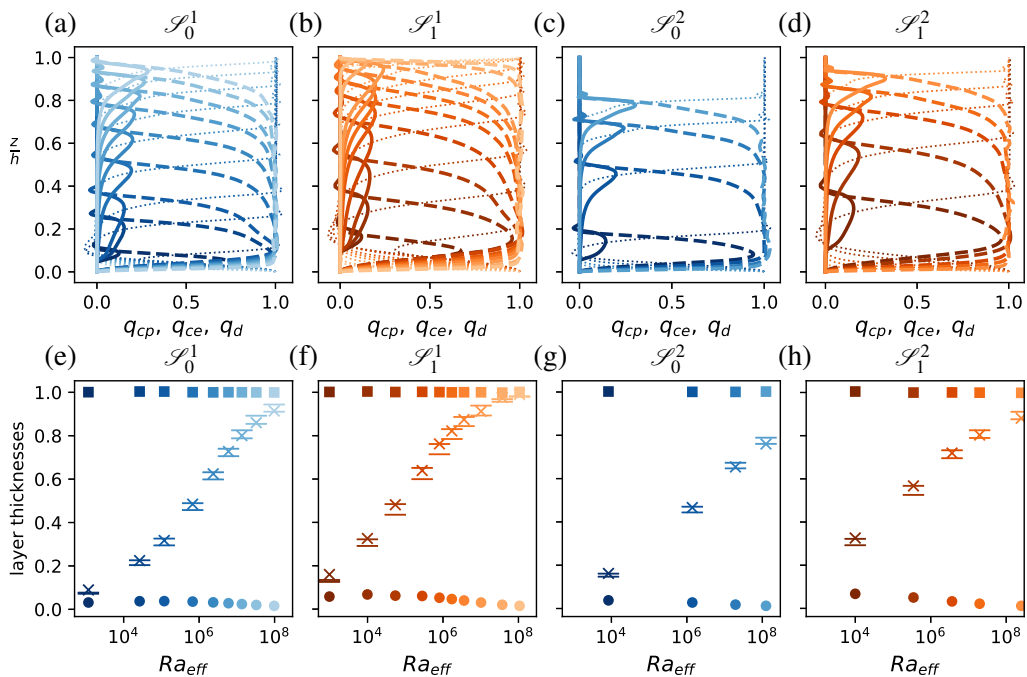
\begin{figure}
\centering
\includegraphics[width=1\textwidth]{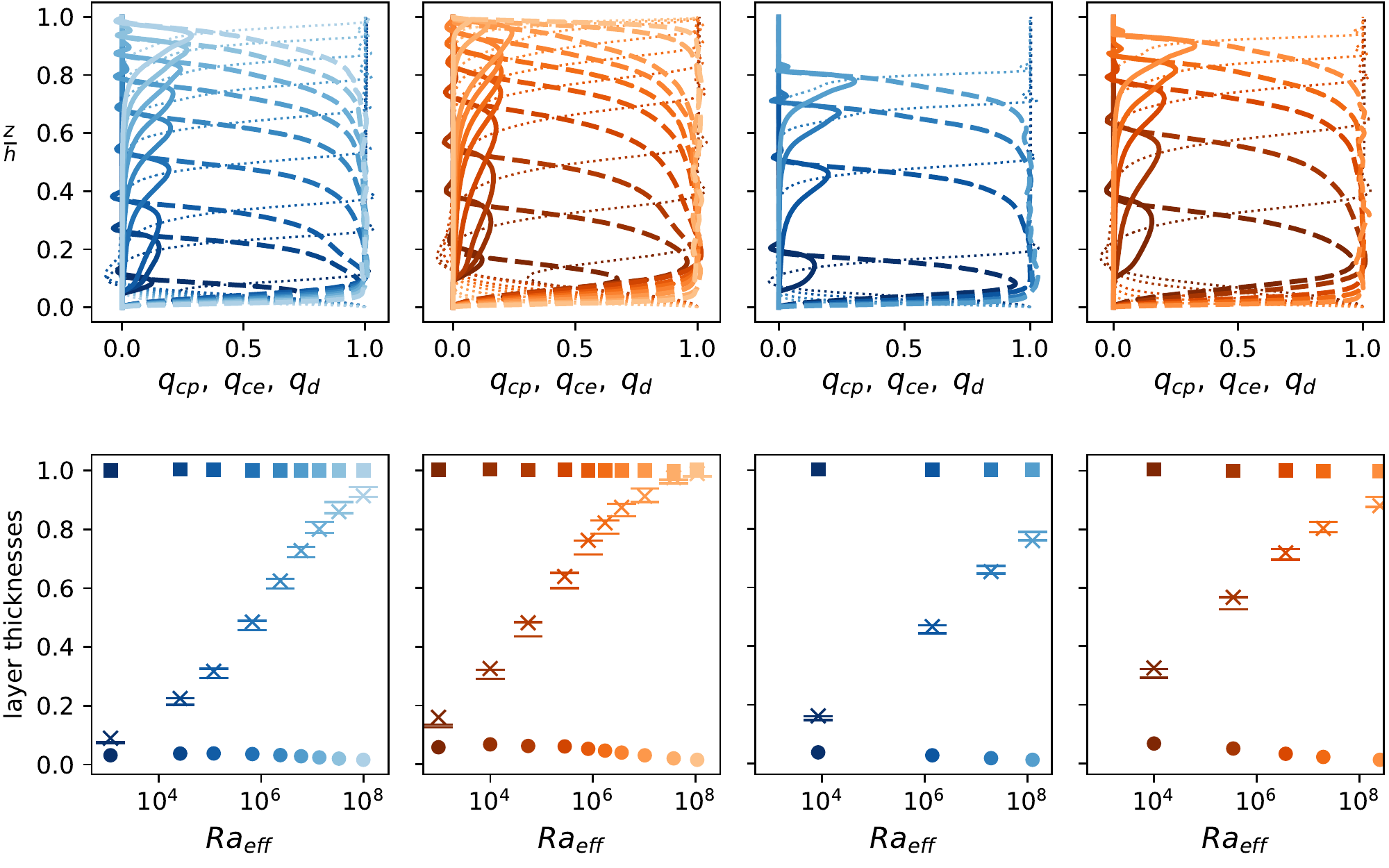}
\put(-370,245){\large{(a)}}\put(-325,245){$\mathcal{S}_0^1$}
\put(-278,245){\large{(b)}}\put(-233,245){$\mathcal{S}_1^1$}
\put(-186,245){\large{(c)}}\put(-142,245){$\mathcal{S}_0^2$}
\put(-94,245){\large{(d)}}\put(-50,245){$\mathcal{S}_1^2$}
\put(-370,120){\large{(e)}}\put(-325,120){$\mathcal{S}_0^1$}
\put(-278,120){\large{(f)}}\put(-233,120){$\mathcal{S}_1^1$}
\put(-186,120){\large{(g)}}\put(-142,120){$\mathcal{S}_0^2$}
\put(-94,120){\large{(h)}}\put(-50,120){$\mathcal{S}_1^2$}
\vspace{-0.05in}\caption{(a)-(d) Convective plume-driven $q_{cp}$ (dashed lines), entrainment-driven $q_{ce}$ (solid lines) and conductive $q_d$ (thin dotted lines) heat fluxes as functions of water depth $z/h$ ($y$ axis) for simulations $\mathcal{S}_0^1$, $\mathcal{S}_1^1$, $\mathcal{S}_0^2$ and $\mathcal{S}_1^2$, respectively. (e)-(h) Top of the bottom conductive layer $\widetilde{z}=L_{db}$ (big bottom circles), of the plume-driven convective layer $\widetilde{z}=L_{db}+L_{cp}$ (lower horizontal bars), of the entrainment-driven convective layer $\widetilde{z}=L_{db}+L_{cp}+L_{ce}$ (upper horizontal bars) and of the stably-stratified layer $\widetilde{z}=L_{db}+L_{cp}+L_{ce}+L_{dt}$ (top squares) as functions of $Ra_{eff}$ for simulations $\mathcal{S}_0^1$, $\mathcal{S}_1^1$, $\mathcal{S}_0^2$ and $\mathcal{S}_1^2$, respectively. The crosses show the height $\widetilde{z}_d$ of the mean $\widetilde{T}_d$ isotherm. The lines' and symbols' color becomes lighter as the heat flux or water depth increases. }
\label{fig8}
\end{figure}

{The lowering of the bulk temperature is caused by turbulent plumes that occasionally penetrate into the stable layer and entrain some of the top cold fluid into the convection zone. The significance of the entrained stable fluid to the system's overall dynamics can be estimated from its contribution to the convective heat flux \cite[][]{Couston2017}. Indeed, the downward motion of cold fluid can contribute a positive heat flux similar to upward-moving warm plumes.} We define the depth-dependent (dimensionless) heat flux due to convective plumes, which we denote by $q_{cp}$, and the depth-dependent heat flux due to the entrained stable fluid, which we denote by $q_{ce}$, as
\ba{}\label{eq:qcs}
q_{cp} = \langle \widetilde{w}\lp \widetilde{T}-\widetilde{T}_d \rp \lp \widetilde{T}>\widetilde{T}_d \rp \rangle, \\
q_{ce} = \langle \widetilde{w}\lp \widetilde{T}-\widetilde{T}_d \rp \lp \widetilde{T}\leq\widetilde{T}_d \rp \rangle,
\ea
respectively. We show in figure \ref{fig8}(a)-(d) the entrained heat flux $q_{ce}$ (solid lines), the plume-driven convective heat flux $q_{cp}$ (dashed lines) and the conductive heat flux $q_d=-\langle \p_{\widetilde{z}} \widetilde{T} \rangle$ (dotted lines) as functions of the normalized water depth $z/h$ for all simulation stages including a stable layer. The plume-driven convective heat flux is significant over a depth that becomes larger with each simulation stage (later stages are shown with lighter colors), due to the increase of either $F$ (figures \ref{fig8}(a)-(b)) or $h$ (figures \ref{fig8}(c)-(d)). The top of the convective layer is approximately where the entrained heat flux is maximum and where the conductive heat flux increases rapidly with $z/h$. Note that $q_{ce}+q_{cp}+q_d\approx 1$ for all $z/h$ as the total heat flux is depth invariant at statistical steady state (the discrepancy after temporal and horizontal averaging is smaller than a few percent). {We remark that the conductive heat flux $q_d$ can be negative in the lower half of the convective layer (see, e.g., dotted lines to the left of the 0 abscissa below $z/h\approx 0.2$ in figures \ref{fig8}(a),(b)) and that the entrainment heat flux $q_{ce}$ can be slightly negative at the top of the convection zone (see, e.g., solid lines going to the left of the 0 abscissa at $z/h\approx 0.2, 0.55, 0.7, 0.8$ in figure \ref{fig8}(c)). The reversals of the conductive heat flux in the convective region and of the convective heat flux at the bottom of the stable layer are real physical phenomena (i.e., they are neither a numerical nor a statistical artefact) and we note that they have been observed in past laboratory experiments \cite[][]{Townsend1964,Adrian1975}.}

\begin{figure}
\centering
\includegraphics[width=0.8\textwidth]{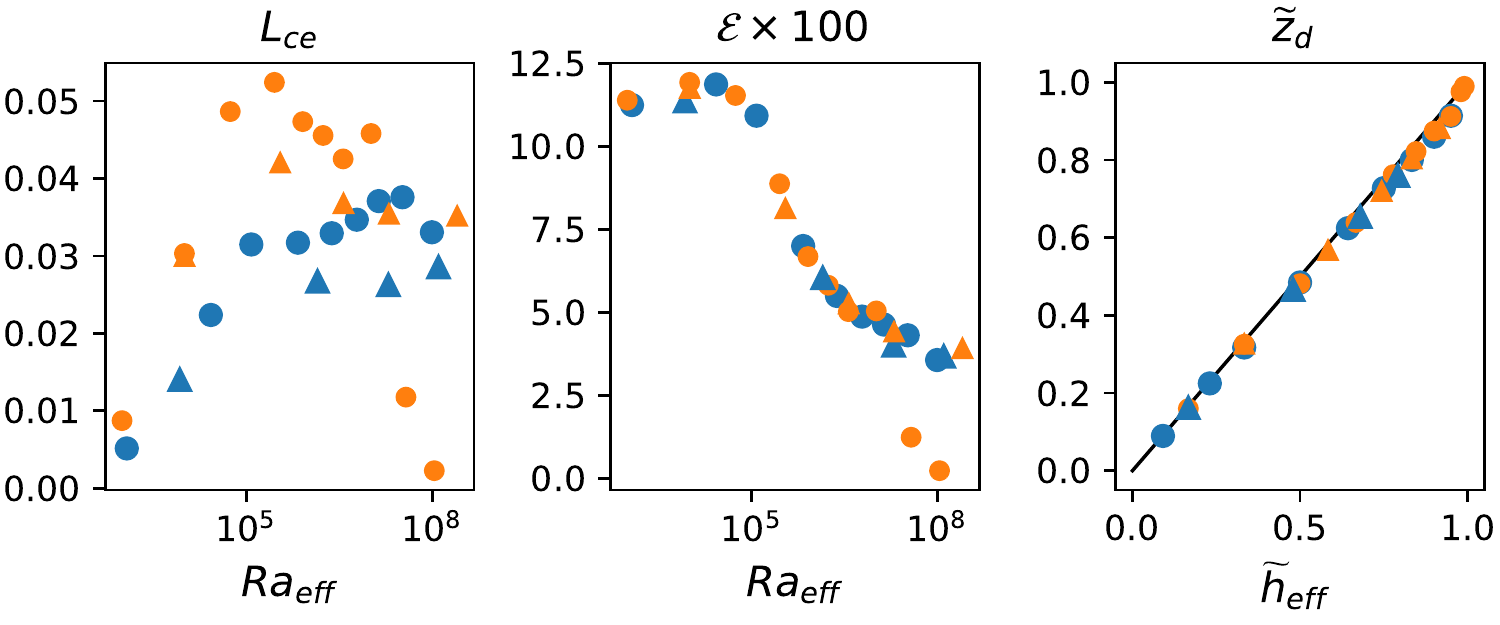}
\put(-300,122){\large{(a)}}
\put(-197,122){\large{(b)}}
\put(-95,122){\large{(c)}}
\vspace{-0.05in}\caption{(a) Thickness of the entrained layer $L_{ce}$ (cf. equation \eqref{eq:eqlthick}) as a function of $Ra_{eff}$. (b) Entrainment parameter $\mathcal{E}$ as a function of $Ra_{eff}$. (c) Dimensionless height $\widetilde{z}_d$ of the $\widetilde{T}_d$ isotherm as a function of the expected convective layer thickness $\widetilde{h}_{eff}$. The solid line shows $\widetilde{z}_d=\widetilde{h}_{eff}$.}
\label{fig9}
\end{figure}

{We estimate the contribution of each heat flux component (i.e., $q_{cp}$, $q_{ce}$ and $q_d$) to the full heat transport by computing the ratio of each heat flux component (integrated over depth) to the volume-averaged heat flux, which is unity in dimensionless space. We separate the contribution of the diffusive heat flux below and above the convection zone by integrating $q_d$ either over $[0,\widetilde{h}_{eff}]$ or over $[\widetilde{h}_{eff},1]$. As in \cite[][]{Couston2017}, we interpret the ratio of each depth-integrated heat flux component to the mean heat flux as an equivalent (dimensionless) layer thickness (wherein the full heat flux is assumed transported by that heat flux component), in order to visualize the contribution of each heat flux component to the full heat transport.} Thus, we denote by 
\ba{}\label{eq:eqlthick}
L_{db} = \int_0^{\widetilde{h}_{eff}}q_d dz, \quad L_{cp} = \int_0^{1}q_{cp} dz, \quad L_{ce} = \int_0^{1}q_{ce} dz, \quad L_{dt} = \int_{\widetilde{h}_{eff}}^{1}q_d dz,
\ea
the equivalent layer thicknesses of the diffusive bottom boundary layer, of the plume-driven convective layer, of the entrained layer and of the diffusive top stable layer, respectively, {which we envision as stacked on top of each other,} with $L_{db}+L_{cp}+L_{ce}+L_{dt}\approx 1$. Figures \ref{fig8}(e)-(h) show the top of each layer, i.e., $\widetilde{z}=L_{db}$ (bottom circles), $\widetilde{z}=L_{db}+L_{cp}$ (lower horizontal bars), $\widetilde{z}=L_{db}+L_{cp}+L_{ce}$ (upper horizontal bars) and $\widetilde{z}=L_{db}+L_{cp}+L_{ce}+L_{dt}$ (top squares) as functions of $Ra_{eff}$. We also display the height $\widetilde{z}_d$ of the mean $\widetilde{T}_d$ isotherm for each simulation (crosses), which is almost always between the top of the plume-driven layer and the top of the entrained layer. Importantly, the thickness of the entrained layer, which is given by the distance between the two horizontal bars does not vary significantly with $Ra_{eff}$. This can be clearly seen in figure \ref{fig9}(a), which shows $L_{ce}$ as a function of $Ra_{eff}$. $L_{ce}$ is typically less then 0.05 and does not noticeably increase with $Ra_{eff}$. While the thickness of the entrained layer remains approximately constant, the thickness of the plume-driven convective layer thickness increases with $Ra_{eff}$, such that the entrainment parameter, $\mathcal{E}$, which we define as
\ba{}\label{eq:ent}
\mathcal{E} = \f{q_{ce}}{q_{cp}+q_{ce}},
\ea
and which compares the entrainment heat flux to the full heat flux, is (almost) monotonically decreasing with $Ra_{eff}$ (figure \ref{fig9}(b)). {Figure \ref{fig9}(c) shows the height $\widetilde{z}_d$ of the $\widetilde{T}_d$ isotherm, which may be considered as an output mean convective--stably-stratified interface height (similar to the top of the entrained layer $\widetilde{z}=L_{db}+L_{cp}+L_{ce}$), as a function of the expected convective layer depth $\widetilde{h}_{eff}=1/(1+\overline{S}(\overline{S}>0))$. We have $\widetilde{z}_d\approx \widetilde{h}_{eff}$ for all simulations and $\widetilde{z}_d\approx L_{db}+L_{cp}\approx L_{db}+L_{cp}+L_{ce}$ since $L_{ce}$ is small (cf. figures \ref{fig8}(e)-(h)), which means that the convective and stably-stratified layers are sufficiently decoupled that $\widetilde{h}_{eff}$ is an accurate estimate for the thickness of the well-mixed turbulent bulk. Note that $\widetilde{h}_{eff}$, $\widetilde{z}_d$ and $L_{db}+L_{cp}+L_{ce}$ are not always close to each other, as is the case here, but can be significantly different when there is strong entrainment \cite[][]{Couston2017}. }

The conclusions of this section are that (i) the entrainment of stable fluid can modify the temperature of the well-mixed bulk in subglacial lakes but that (ii) the influence of the stably-stratified layer on the convective dynamics does not increase with the effective Rayleigh number or heat flux and water depth. Thus, we predict that the convective and stably-stratified layers can be considered decoupled at leading order and that classical scaling laws for $Nu$ and $Re$ apply to natural subglacial lakes with large water depths and heat fluxes, provided that the effective Rayleigh number in equation \eqref{eq:Raeff} is considered. {Further, we remark that the (finite) lowering of the bulk temperature from $(\widetilde{T}_{bulk}-\widetilde{T}_{d})\approx (\langle \widetilde{T}_{b}\rangle-\widetilde{T}_{d})/2$ to $(\widetilde{T}_{bulk}-\widetilde{T}_{d})\approx (\langle \widetilde{T}_{b}\rangle-\widetilde{T}_{d})/4$ in the presence of a thick stable layer can be accounted for by changing the effective temperature difference driving the convection from $\widetilde{\Delta}_{eff}$ to $3\widetilde{\Delta}_{eff}/2$. Multiplying the Nusselt number by 2/3 and the effective Rayleigh number by 3/2 for the simulation results with a stable layer indeed provides an improved collapse with the simulation results without a stable layer (cf. light-colored symbols in figures \ref{fig6}(a)-(c)).} {We note that the lowering of the normalized bulk temperature in the presence of a stable layer, although relatively constant over the range of simulations considered (cf. figure \ref{fig7}(b)), should be expected (in general) to vary with the problem parameters \cite[as is the case in][]{Couston2017}, such that the 3/2 correction proposed for the effective temperature difference is unlikely to be universal. Investigating the functional form of the correction coefficient for the effective temperature difference with the problem parameters in the context of real subglacial lakes would be interesting but is beyond the scope of the present work and is not expected to significantly change the current leading-order predictions, which neglect the influence of the stable layer on the convection.}

\section{Geophysical discussion}\label{sec:discussion}

\begin{figure}
\centering
\includegraphics[width=1\textwidth]{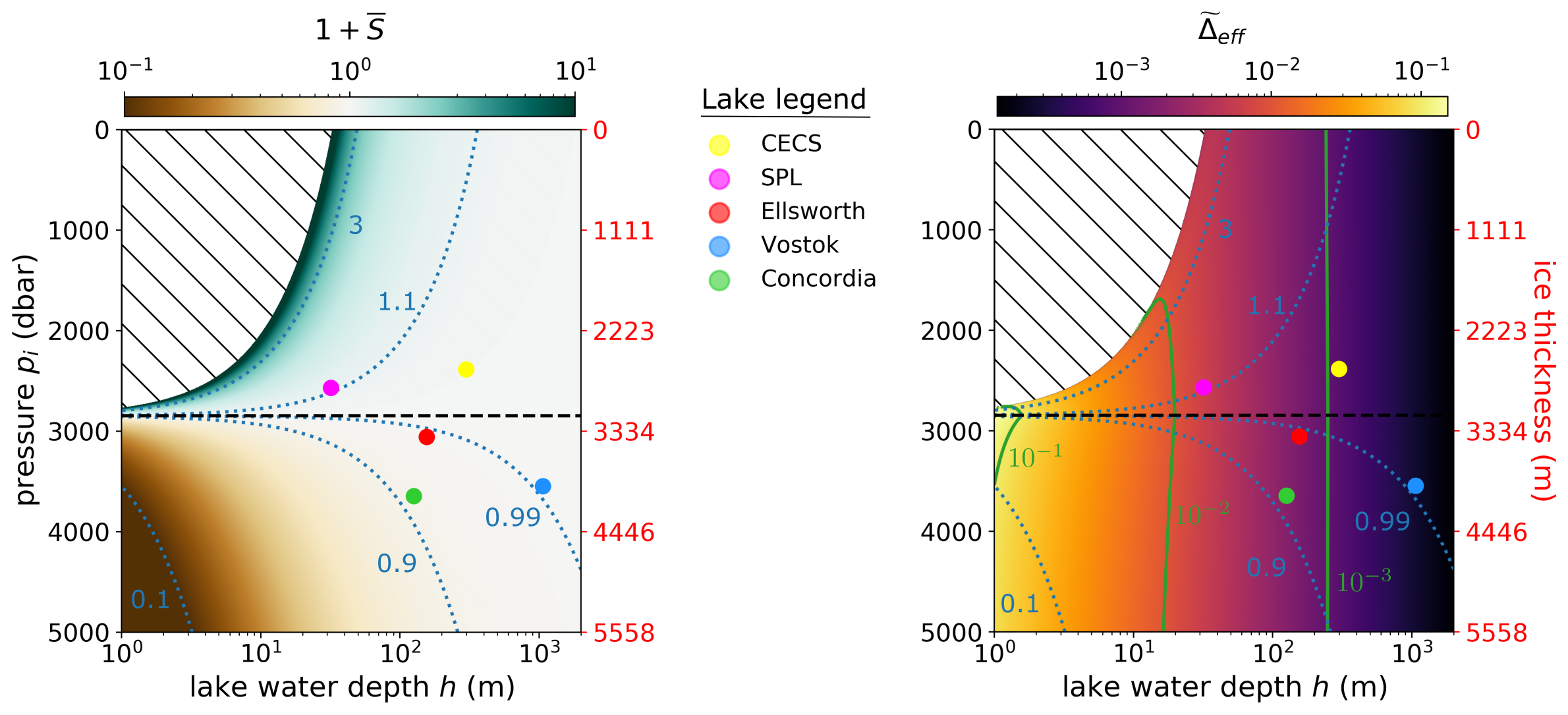}
\put(-377,160){\large{(a)}}
\put(-165,160){\large{(b)}}
\vspace{-0.05in}\caption{(a) Color map of $1+\overline{S}$, with $\overline{S}$ the base-state stiffness parameter, in $(p_i,h)$ space relevant to Antarctic subglacial lakes with heat flux $F=68$ mW/m$^2$. The hashed region in the top left corner highlights subglacial lakes that are fully stable because $F=68$ mW/m$^2$ is smaller than the threshold heat flux $F_t$. The horizontal dashed line highlights the $p_*$ isobar and corresponds also to the isocontour $1+\overline{S}=1$. The dotted blue lines are isocontours of $1+\overline{S}$ and the colored circles (see legend) highlight the location of some of the well-known subglacial lakes in parameter space \cite[][]{Couston2020}. (b) same as (a) but for the dimensionless effective thermal driving $\widetilde{\Delta}_{eff}$. The dotted blue lines still show the isocontours of $1+\overline{S}$ whereas the green solid lines show isocontours of $\widetilde{\Delta}_{eff}$.}
\label{fig10}
\end{figure}

The closed-form expressions presented in table \ref{tab:scalings} are applicable to subglacial lakes that are either in the OB or LEC regime. We recall that the criterion for the OB regime (cf. equation \eqref{eq:der2}) is $\widetilde{\beta}_{eff}\approx 1$, i.e., $\overline{S}\approx -1$, which can be satisfied by fully-convective lakes only. The criterion for the LEC regime is instead $\widetilde{\beta}_{eff}\approx \widetilde{\Delta}_{eff}(1+\overline{S})/2$, i.e., $\widetilde{\Delta}_{eff} \gg 2\overline{S}(\widetilde{T}_d<0)/(1+\overline{S})$, which is satisfied when $\overline{S}\geq 0$, i.e., all lakes with a top stable layer, or $\widetilde{\Delta}_{eff} \gg |\overline{S}| \approx 0$, which may be verified for some fully-convective lakes. We show in figure \ref{fig10}(a) $(1+\overline{S})$ as a function of ice pressure and lake water depth. We consider a heat flux of $F=68$ mW/m$^2$, which is expected to be the average geothermal flux across Antarctica \cite[][]{Martos2017}. For $1+\overline{S}<0$, $\overline{T}_b<T_d$, such that the thermal expansion coefficient is negative throughout the full water column and the lake is stable. {We have} $1+\overline{S}<0$ where the water depth $h$ and ice overburden pressure $p_i$ are small (hashed region in the top left corner), so small that $F=68$ mW/m$^2$ is smaller than the threshold heat flux $F_t(p_i,h)$ (cf. equation \eqref{eq:threshold}). We have $1+\overline{S} \geq 1$ for all lakes with $p_i\leq p_*$ (as expected) and $1+\overline{S} \geq 0.9$ for many fully-convective lakes, including the three well-known subglacial lakes Ellsworth, Concordia and Vostok (cf. three filled circles below the $p_*$ isobar). This means that the LEC regime is applicable to all mixed convective and stably-stratified lakes but also possibly to some of the fully-convective lakes if $\widetilde{\Delta}_{eff}$ is not too small. We have $1+\overline{S}\leq 0.1$, which is the criterion for the OB regime, for a limited number of fully-convective lakes only (bottom left corner below the 0.1 isoline of the diagram), because all lakes that are deep can experience relatively large temperature differences, such that the thermal expansion coefficient becomes variable.

In order to investigate the applicability of the LEC regime to real subglacial lakes, we display in figure \ref{fig10}(b) the dimensionless effective temperature difference $\widetilde{\Delta}_{eff}$ predicted based on equation \eqref{eq:der5} as a function of $(p_i,h)$ and superimpose isocontours of $1+\overline{S}$ (same blue dotted lines as seen in figure \ref{fig10}(a)). We find that $\widetilde{\Delta}_{eff}$ is always smaller or much smaller than $|\overline{S}|$, i.e., such that the LEC regime is not applicable unless $\overline{S} \geq 0$, except for a narrow band of ice pressures $p_i$ close to $p_*$ (horizontal dashed line). For instance, the isoline $\widetilde{\Delta}_{eff}=10^{-2}$ is the rightmost boundary of the region where $\widetilde{\Delta}_{eff}\geq 10^{-2}$, and it can be seen that this region does not overlap with the region of $|\overline{S}|\leq 10^{-2}$, which is above and to the right of the $1+\overline{S}=0.99$ isocontour, unless $\overline{S}\geq 0$ (or $p_i\leq p_*$). Thus, the $-\overline{S}(\widetilde{T}_d<0)$ term in equation \eqref{eq:der2} is not negligible when $\widetilde{T}_d<0$ (or $-1\leq \overline{S}<0$) and neither the OB regime nor the LEC regime is applicable to deep fully-convective lakes.

Although neither limiting regime is rigorously applicable to deep fully-convective lakes, we can still use the closed-form expressions in table \ref{tab:scalings} in order to provide some first-order estimates for the rms velocity and for the temperature difference driving the convection in both shallow subglacial lake CECs, where the LEC regime applies, and deep subglacial lake Vostok. The most recent geophysical information on lake CECs is that the heat flux should be around $F\approx 100$ mW/m$^2$ and that the maximum water depth is $h\approx 300$ m\footnote{Personal communication with Nicolás Donoso from Centro de Estudios Científicos, Chile.}. Using $F=100$ mW/m$^2$, $h=300$ m and using the approximate pre factors and exponents $a= 0.25$, $c= 0.006$, $\alpha = 2/7$ and $\gamma = 3/5$ for the closed-form expressions in the LEC regime yields $V_{rms}\approx 2.4$ mm/s and $\Delta_{eff} \approx 0.04 \C$, which corresponds to a bottom temperature $T_b = T_d + \Delta_{eff} = -1.12\C$, i.e., $0.74\C$ above freezing ($T_f=-1.86\C$). For lake Vostok, the geothermal flux is closer to $50$ mW/m$^2$ and the maximum water depth is on the order of 1000 m \cite[][]{Siegert2001a}. Using $F=50$ mW/m$^2$, $h=1000$ m and $a= 0.25$, $c= 0.006$, $\alpha = 2/7$ and $\gamma = 3/5$ yields in the LEC regime $V_{rms}\approx 4.2$ mm/s and $\Delta_{eff} \approx 0.03 \C$, with $\Delta_{eff}$ directly equal to $T_b - T_f$, i.e., the bottom temperature in excess of the freezing point ($T_f=-2.83 \C$), since there is no stable layer in this case. Note that we obtain $\Delta_{eff} \approx 0.005 \C$ and $V_{rms}\approx 25$ mm/s, i.e., a significantly larger rms velocity, using predictions in the OB regime for lake Vostok because the effective thermal expansion coefficient is strongly overestimated in the OB regime for deep fully-convective lakes. 

The velocities in the LEC regime are of the same order as the velocities $V_{rms}\approx 1$ mm/s for lakes CECs and $V_{rms}\approx 4$ mm/s for lake Vostok predicted by previous studies, including \cite[][]{Wuest2000}, who applied scaling laws of vertical convection in the rapidly-rotating regime to lake Vostok, and, more recently, \cite[][]{Couston2020}, who predicted the physical properties of most subglacial lakes based on three-dimensional non-rotating scaling laws. The rough agreement between the velocities predicted in lake Vostok by \cite[][]{Wuest2000} and this work is far from trivial since the former assumes that the flow dynamics is dominated by rotation while the latter neglects rotation. The importance of rotation in subglacial lakes is an open question, which is beyond the scope of this manuscript. The agreement between the velocities predicted in lakes CECs and Vostok by \cite[][]{Couston2020} and this work is also better than could be expected as there is a significant difference in the scaling exponent $\gamma\approx 3/5$ calculated in this work based on two-dimensional simulations and the scaling exponent $\gamma\approx 1/2$ used in \cite[][]{Couston2020} and obtained based on three-dimensional simulations \cite[][]{King2013}. The discrepancy in exponents is compensated by the difference in pre factors for the case of lakes CECs and Vostok considered here but may result in inaccurate predictions of three-dimensional rms velocities in other conditions. Another difference between this work and \cite[][]{Couston2020} is that the latter study used conservative estimates for the effective thermal expansion coefficient, such that their predicted velocities can be expected to be smaller than those we estimate based on the closed-form expression for the thermal expansion coefficient derived in the LEC regime. We refer the reader to \cite[][]{Couston2020} for a detailed geophysical discussion of first-order predictions of physical properties in most Antarctic subglacial lakes, which is based on the assumption--which we validated in this work through novel direct numerical simulations--that the convective and stably-stratified  dynamics are decoupled at leading order.

We have demonstrated that the penetration of convective motions tends to lower the temperature of the turbulent well-mixed bulk when the top stable layer is not too thin, i.e., much thicker than the thermal boundary layer, despite the fact that the convection and stably-stratified layer can be considered dynamically decoupled at leading order. \cite{Couston2020} showed that the top stable layer of Antarctic subglacial lakes grows from 0 m at $p_i=p_*$ to approximately 40 m thickness at $p_i=p_{atm}\approx 10^5$ Pa. The thickness of the thermal boundary layer can be estimated as $h_{eff}/(2Nu)$ and is less than 1 m for most Antarctic subglacial lakes \cite[][]{Couston2020}. Thus, we expect that most mixed convective and stably-stratified subglacial lakes have bulk temperature that can be lowered by entrainment, but note that the decrease of the bulk temperature in three-dimensional simulations may differ from the lowering we reported in section \S\ref{sec:entrainment} and figure \ref{fig7}(b).

\section{Concluding remarks}\label{sec:conc}

We have investigated the dynamics of laboratory-scale subglacial lakes via direct numerical simulations and demonstrated that the Nusselt number $Nu$ and Reynolds number $Re$ follow similar scalings laws as in classical Rayleigh-B\'enard convection provided that an effective Rayleigh number $Ra_{eff}$ is considered. {We obtained pre-factors and exponents similar to those obtained for two-dimensional fully-convective Rayleigh-B\'enard simulations in the Oberbeck-Boussinesq regime with $Pr=1$ \cite[][]{Johnston2009,Sugiyama2009}, which means that the effect of the Prandtl number is relatively weak.} {As in \cite[][]{Johnston2009}, we remark that it is possible to define an effective flux-based Rayleigh number, i.e., $Ra_{Feff} = ( gh_{eff}^4F\beta_{eff})/\lp k\nu\kappa\rp$, which is equal to $\overline{Ra}_F$ in the OB regime (i.e., when the thermal expansion coefficient is constant) and is related to $Ra_{eff}$ through $Ra_{Feff} = NuRa_{eff}$.}

We have shown that dimensional variables, such as the effective temperature difference, or bottom temperature, and rms velocity, scales differently with the problem parameters depending on whether the effective thermal expansion coefficient $\beta_{eff}$ is constant or linearly proportional to the effective temperature difference $\Delta_{eff}$. We have called the dynamical regimes associated with the two limiting behaviors of $\beta_{eff}$ the OB regime and the LEC regime and derived explicit expressions for all variables of interest in terms of the problem parameters in both regimes (table \ref{tab:scalings}). When $\beta_{eff}$ is an affine function of temperature, with non-neligible $y$-intercept, it is possible to infer $\Delta_{eff}$ from equation \eqref{eq:der3} (and deduce all other variables onward) but an explicit expression is not available. {We remark that the expressions for, e.g., $\Delta_{eff}$, with the problem parameters are discontinuous between the OB regime and the LEC regime but can be continuated and connected through the use of equation \eqref{eq:der3} in regions of the parameter space where neither regime is accurate.}

{The key results of our work are: 
\begin{enumerate}
\item the definition of an accurate $Ra_{eff}$ for subglacial lakes leading to classical scaling laws for $Nu$ and $Re$ with $Ra_{eff}$,
\item the demonstration that the convective and stably-stratified layer dynamics are decoupled at leading order,
\item the identification of the two limiting OB and LEC regimes,
\item the derivation of closed-form expressions for all variables of interest in terms of the problem parameters in the OB and LEC regimes.
\end{enumerate}
}The numerical predictions of physical variables in subglacial lakes is briefly discussed in section \S\ref{sec:discussion}. We emphasize that while we expect that the expressions hold in both two and three dimensions, the pre factors $a$, $c$ and exponents $\alpha$, $\gamma$ may vary between two-dimensional and three-dimensional simulations. Clearly, future predictions of physical variables in subglacial lakes should consider power laws obtained in three dimensions, {although we remark that the power law for $Nu$ is almost the same between two-dimensional and three-dimensional simulations of both classical Rayleigh-B\'enard convection \cite[][]{Ahlers2009} and water convection close to the density maximum \cite[][]{Wang2019}.}

We have demonstrated that the penetration of convective motions in the stratified layer decreases with $Ra_{eff}$, such that there should be limited entrainment in most two-layer subglacial lakes. This does not mean that future studies should discard the stratified layer altogether. For instance, it would be interesting to investigate if internal waves excited by convection \cite[][]{Couston2018b} can melt the ice ceiling such that ice-trapped oxygen and nutrients remain available in subglacial lakes with a thin ice cover. {The independence of the bulk temperature or thickness of the entrained layer with increasing $Ra_{eff}$ or decreasing $S_{eff}$, which we observed for simulations with a thick stable layer, seems at odds with the results of \cite[][]{Couston2017}. \cite{Couston2017} demonstrated a monotonic lowering of the bulk temperature and increase of the entrained layer thickness with a decrease of their input stiffness $S_i$, which, intuitively, is what may be expected if the inverse of the stiffness does provide a measure of the degree of coupling between the convective and stably-stratified layers. We expect that the linear dependence of the thermal expansion coefficient with the temperature in our numerical simulations may be responsible for the observed discrepancy between our results and the results of \cite[][]{Couston2017}, who instead used a piecewise-constant thermal expansion coefficient. On the one hand, in our numerical simulations, plumes approaching the $\widetilde{T}_d$ isotherm lose their buoyancy even before entering the stable layer, which limits penetration. On the other hand, the base of the stable layer always has a small negative thermal expansion coefficient, i.e., a small Brunt-V\"ais\"al\"a frequency, such that penetration is always possible, i.e., including at large stiffness $S_{eff}$. Thus, the stiffness parameter can be expected to have a weaker effect on entrainment when the thermal expansion coefficient changes sign smoothly rather than discontinuously. A study designed specifically to investigate the effect of the functional form of the thermal expansion coefficient with temperature on entrainment would be a valuable addition to the literature on penetrative convection but is beyond the scope of this work.}

Future works could investigate {the effect of planetary rotation, which is most important near the poles and may play a significant role in the turbulent dynamics and mean temperatures obtained in Antarctic subglacial lakes \cite[][]{Wuest2000}, and low to moderate salt concentrations. Considering planetary rotation will require three-dimensional simulations, unless we assume decoupled rotating convective and stably-stratified dynamics and use existing knowledge of planetary rotation effects on classical Rayleigh-B\'enard convection \cite[][]{Plumley2019}.} An important limitation of the present work is the assumption of a flat ice-water interface, such that considering a tilted ice ceiling and investigating the combined dynamics of the resulting baroclinic horizontal flow with the vertical convection is essential. There is also a possibility that the proposed scalings may become inaccurate as the water depth becomes large enough that the effect of pressure variations within the water column on buoyancy can no longer be neglected.

We expect that this paper and future studies improving our understanding of the hydrodynamic conditions in subglacial lakes could help identify subglacial environments that are physically favourable for a biome and guide future observations and sampling of subglacial lake water. 

\appendix

\section{Pressure effects}\label{sec:appA}

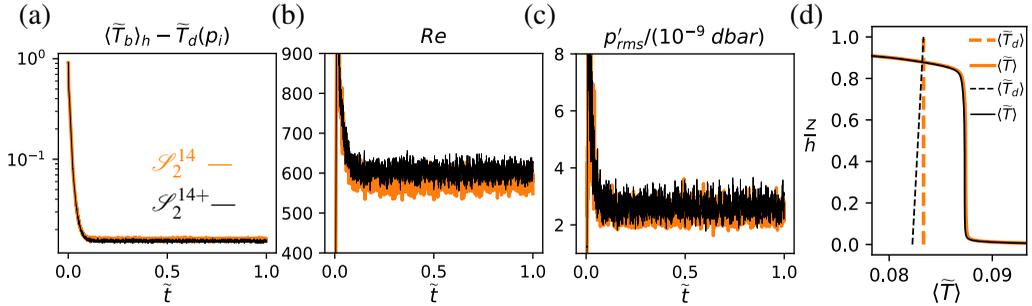
\begin{figure}
\centering
\includegraphics[width=1\textwidth]{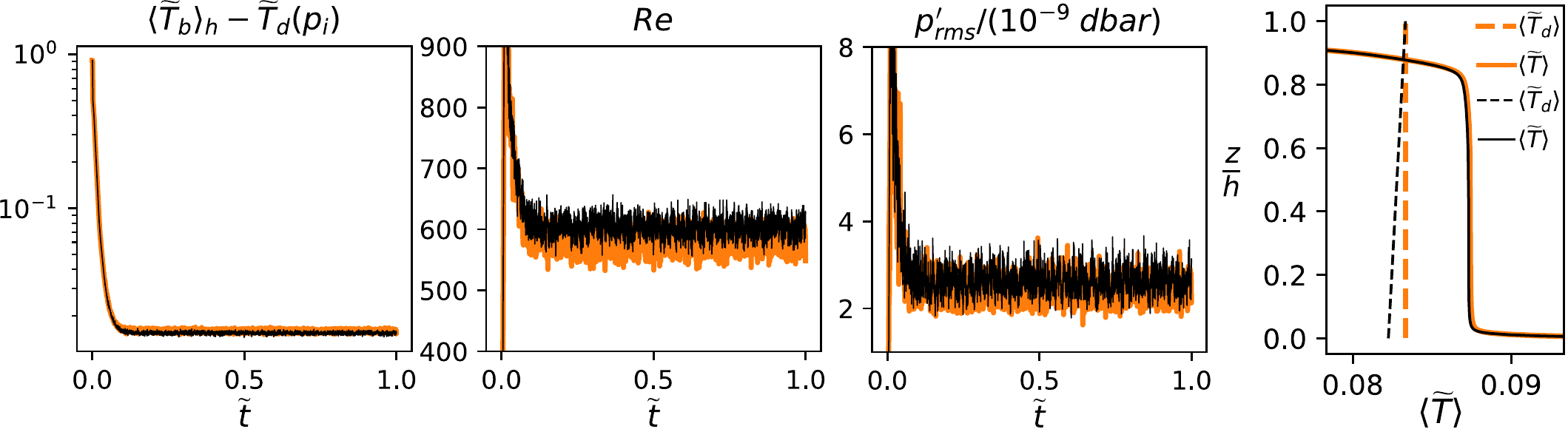}
\put(-380,105){\large{(a)}}
\put(-285,105){\large{(b)}}
\put(-190,105){\large{(c)}}
\put(-90,105){\large{(d)}}
\put(-330,50){\orange{$\mathcal{S}_2^{14}\;$  \Lshort}}
\put(-330,35){$\mathcal{S}_2^{14+}$\Lshort}
\vspace{-0.05in}\caption{(a) Time history of the dimensionless effective temperature difference $\langle\widetilde{T}_b\rangle_h-\widetilde{T}_d(p=p_i)$ for simulation $\mathcal{S}_{14}^2$ (thick orange line), which is simulation $\mathcal{S}_{1}^2$ discussed in the main text with $h = 4$ m and constant $\widetilde{T}_d$, and simulation $\mathcal{S}_{14+}^2$ (thin black line), which includes hydrostatic effects in $\widetilde{T}_d$. (b) same as (a) but for the Reynolds number. (c) same as (a),(b) but for the dynamic pressure rms $p'_{rms}$ normalized by $10^{-9}$ dbar. (d) Mean vertical profiles of the dimensionless temperature $\widetilde{T}$ (solid lines) and of the dimensionless temperature of maximum density $\widetilde{T}_d$ (dashed lines) for simulations $\mathcal{S}_{14}^2$ (thick orange lines) and  $\mathcal{S}_{14+}^2$ (thin black lines).}
\label{fig11}
\end{figure}

All simulations discussed in the main text and listed in table \ref{tab:sims} assume that the thermal expansion coefficient can be approximated as $\beta(p,T)\approx\beta(p_i,T)$, i.e., such that pressure variations within the water column are neglected in the expression for $\beta$. We demonstrate that this is a valid approximation at leading order by showing in figure \ref{fig11} the time history and vertical profiles of several variables obtained in simulation $\mathcal{S}_{14}^2$, which is simulation $\mathcal{S}_{1}^2$ with $h=4$ m and in an additional simulation, denoted  $\mathcal{S}_{14+}^2$, which is the same as $\mathcal{S}_{14}^2$ but with $\beta(p,T)\approx\beta(p_i+\rho_0g(h-z),T)$, i.e., such that it includes  pressure variations due to hydrostasy in the expression of the thermal expansion coefficient. We select $\mathcal{S}_{14}^2$ as a point of comparison as the temperature of maximum density, $T_d$, which is the variable that is most sensitive to $p$ in the expression for $\beta$, is attained inside the water column (i.e., there is a top stable layer) and because it is one of the simulations with the largest water depth (4 meters), such that pressure variations due to hydrostasy are relatively large.

Figures \ref{fig11}(a)-(c) show that the dimensionless effective temperature difference, the Reynolds number, and the dynamic pressure rms $p'_{rms} =[\langle \lp p - \langle p \rangle_h \rp^2 \rangle_h]^{1/2}$, which we evaluate in (approximately) the middle of the convection zone at $z=0.5$, are similar in both cases. The small discrepancies between the two cases are that the dimensionless effective temperature difference is slightly smaller in $\mathcal{S}_{14+}^2$ than in $\mathcal{S}_{14}^2$ and that $Re$ and $p'_{rms}$ are slightly larger in $\mathcal{S}_{14+}^2$ than in $\mathcal{S}_{14}^2$. This suggests that $\mathcal{S}_{14+}^2$ is slightly more turbulent than $\mathcal{S}_{14}^2$, which is {expected} since the decrease of $T_d$ with decreasing $z$ in $\mathcal{S}_{14+}^2$ results in a slightly larger thermal expansion coefficient in $\mathcal{S}_{14+}^2$ than in $\mathcal{S}_{14}^2$ at depth. The variability of $T_d$ within the water column for $\mathcal{S}_{14+}^2$ can be seen in figure \ref{fig11}(d), which displays the mean vertical profiles of $\widetilde{T}_d$ and $\widetilde{T}$. It can be seen that the decrease of $\langle\widetilde{T}_d\rangle$ with depth in $\mathcal{S}_{14+}^2$ is significantly smaller than the variability of $\langle\widetilde{T}\rangle$, which is, ultimately, the reason why the neglect of pressure variations in the expression for the thermal expansion coefficient is valid, at least for the laboratory-scale subglacial lakes considered in this paper. For deeper (geophysical) subglacial lakes, it may be expected that the decrease of $\widetilde{T}_d$ with depth results in slightly more turbulent conditions than could be predicted assuming $\beta(p,T)\approx\beta(p_i,T)$. The exact discrepancy due to the neglect or consideration of hydrostatic pressure variations in the expression for $\beta$ in geophysical lakes is beyond the scope of this work. We note that considering hydrostatic effects in the expression for $\beta$ does not incur any computational overhead, such that they should be included in future works. Figure \ref{fig11}(c) shows that the dynamic pressure rms is much smaller than 1 dbar (by 9 orders of magnitude), which is equal to the change of hydrostatic pressure over 1 meter. Thus, dynamic pressure variations can be safely neglected from the expression for $\beta$ in simulations of laboratory-scale subglacial lakes as well as in simulations of (most) geophysical subglacial lakes.

\section*{Acknowledgements}

I gratefully acknowledge fruitful discussions with Benjamin Favier and Thierry Alboussière as well as many constructive comments from four anonymous reviewers, which helped me significantly extend and improve the initial manuscript. This project has received funding from the European Union's Horizon 2020 research and innovation programme under the Marie Sklodowska-Curie grant agreement 793450. I acknowledge PRACE for awarding me access to Marconi at CINECA, Italy.

\textbf{Declaration of Interests.} The authors report no conflict of interest.

\vspace{-0.1in}
\bibliographystyle{jfm}
\bibliography{subglaciallakes}

\end{document}